\documentclass[traditabstract]{aa}  
\usepackage{natbib}
\bibpunct{(}{)}{;}{a}{}{,}
\usepackage{amssymb}
\usepackage{longtable}
\usepackage{txfonts}
\usepackage{graphicx}
\usepackage{pdflscape}
\newcommand{\iso}[1]{\mbox{$^{#1}{\rm Ba}$}}
\newcommand{\eps}[1]{\log\varepsilon_{\rm #1}}

\newcommand{\Teff}{T$_{\rm eff}$~} 
\newcommand{\Teffno}{T$_{\rm eff}$} 

\newcommand{\logg}{\ensuremath{\log g}~}
\newcommand{\loggno}{\ensuremath{\log g}}

\newcommand{\vmic}{$v_\mathrm{t}$}
\newcommand{\msol}{M$_\odot$~}

\newcommand{\kms}{\,km\,s$^{-1}$}         

\begin{document} 

   \title{The early days of the Sculptor dwarf spheroidal galaxy\thanks{Based on ESO programs 087.D-0928(A) and  091.D-0912(A)}}


   \author{P. Jablonka
          \inst{1,2}
          \and
          P. North \inst{1}
          \and
          L. Mashonkina \inst{3}
          \and
          V. Hill   \inst{4}
          \and
          Y. Revaz \inst{1}
          \and
          M. Shetrone \inst{5}
          \and 
           E. Starkenburg \inst{ 6,7, \thanks{CIFAR Global Scholar  }}
          \and
          M. Irwin \inst{8}
          \and       
          E. Tolstoy \inst{9}
          \and 
          G. Battaglia \inst{10,11}
          \and
          K. Venn \inst{12}          
          \and
          A. Helmi \inst{9}    
          \and
          F. Primas \inst{13} 
          \and 
          P. Fran\c cois\inst{2}
                             }

   \institute{
Laboratoire d'astrophysique, Ecole Polytechnique F\'ed\'erale de Lausanne (EPFL),  Observatoire, CH-1290 Versoix, Switzerland
              \email{pascale.jablonka@epfl.ch }
\and   GEPI, Observatoire de Paris, CNRS, Universit\'e de Paris Diderot, F-92195  Meudon, Cedex, France
\and   Institute of Astronomy, Russian Academy of Sciences, RU-119017 Moscow,  Russia
\and   Department Cassiopee, University of Nice Sophia-Antipolis, Observatoire de C\^{o}te d'Azur, CNRS, F-06304 Nice Cedex 4, France
\and   McDonald Observatory, University of Texas, Fort Davis, TX 79734, USA
\and   Department of Physics and Astronomy, University of Victoria, PO Box 3055 STN CSC, Victoria, BC V8W 3P6, Canada
\and   Leibniz-Institut für Astrophysik Potsdam, An der Sternwarte 16, 14482 Potsdam, Germany
\and   Institute of Astronomy, University of Cambridge, Madingley Road, Cambridge CB3 0HA, UK
\and   Kapteyn Astronomical Institute, University of Groningen, PO Box 800, 9700AV Groningen, the Netherlands
\and   Instituto de Astrofisica de Canarias, calle via Lactea s/n, 38205 San Cristobal de La Laguna (Tenerife), Spain
\and   Universidad de La Laguna, Dpto. Astrofísica, E-38206 La Laguna, Tenerife, Spain
\and   Department of Physics and Astronomy, University of Victoria, 3800 Finnerty Road, Victoria, BC, V8P 1A1, Canada
\and   European Southern Observatory, Karl-Schwarzschild-str. 2, D-85748, Garching bei M\"{u}nchen, Germany
             }

   \date{Received early; accepted soon}

\abstract{ We present the high-resolution spectroscopic study of five
    $-3.9\leq$[Fe/H]$\leq-2.5$ stars in the Local Group dwarf spheroidal,
    Sculptor, thereby doubling the number of stars with comparable observations in
    this metallicity range. We carry out a detailed analysis of the chemical
    abundances of $\alpha$, iron peak, and light and heavy elements, and draw
    comparisons with the Milky Way halo and the ultra-faint dwarf stellar
    populations.  We show that the bulk of the Sculptor metal-poor stars follow
    the same trends in abundance ratios versus metallicity as the Milky Way stars.
    This suggests similar early conditions of star formation and a high degree of
    homogeneity of the interstellar medium. We find an outlier to this main
    regime, which seems to miss the products of the most massive of the Type II
    supernovae. In addition to its  help in refining galaxy formation models,
    this star provides clues to the production of cobalt and zinc. Two of our
    sample stars have low odd-to-even barium isotope abundance ratios, suggestive
    of a fair proportion of s-process. We discuss the implication for the
    nucleosynthetic origin of the neutron capture elements. }


   \maketitle
%

\section{Introduction}

Star formation in dwarf galaxies has been the focus of many recent galaxy
formation simulations.  An extremely wide  variety of topics are affected by the
processes at play, from the evolutionary core/cusp-shape of the dark matter
density profiles \cite[e.g.][]{Teyssier2013}; to the ``too big to fail problem'',
which raises questions about the matching between dark matter halos and their stellar
masses \cite[e.g.][]{Boylan-Kolchin2011, Sawala2013}; to the nature of dark
matter \cite[e.g.][]{governato2015}; and to the identification of the sources
that were able to reionize the universe \cite[e.g.][]{Wise2014}.  Understanding
such processes is then fundamentally necessary in order to accurately simulate
these processes.

 Although still limited, spectroscopic surveys of dwarf spheroidal galaxies (dSphs)
 have already shed substantial light on their evolution.  It follows that their star
 formation efficiency is much lower than that of the Milky Way (MW), as revealed by
 the comparisons between the chemical imprints of dSphs and the MW.  Indeed, the
 metallicity ([Fe/H]) at which Type Ia supernovae (SNeIa) dominate the chemical
 evolution of these small systems is at least a dex smaller than in the MW
 \citep{koch2008,THT09,letarte2010, kirby2011, lemasle2014, hendricks2014}.  Detailed
 abundances therefore allow us to place restrictions on the mass range and the period
 during which small galactic systems could have merged to form larger ones, as
 they need to share similar chemical patterns.  Another merit of investigating the
 elemental abundances of individual stars inside dSphs resides in their power to help
 in the identification of stellar nucleosynthesis sites.  The very different
 evolutionary paths of dSphs result in very distinct chemical signatures providing a
 series of constraints to the models.  For instance, this is the case of the
 neutron-capture elements.  The gradual enrichment in r-process elements may well
 depend on the galactic baryonic mass.  Below [Fe/H] $\sim -3.5$ all galaxies seem to
 have similar, very low, levels of barium and strontium.  At higher metallicities,
 the smallest dwarfs stay at this low level, whereas more massive galaxies such as
 Sextans dSph eventually reach the solar value observed in the MW
 \citep{tafel2010}.This type of evidence is crucial to helping distinguish among
 various possible origins such as a specific type of core collapse supernovae
 \citep[e.g.][]{winteler2012}, neutron star mergers \citep[e.g.][]{wanajo2014}, or
 spinstars \citep[e.g.][]{cescutti2013}. Another example is provided by the
 carbon. Very few carbon-rich extremely metal-poor stars have been found in dSphs, in
 contrast with the Galactic halo \citep{skuladottir2015}.  This differential
 signature provides pieces of evidence to be interpreted on the type of stars whose
 nucleosynthetic production can be retained or accumulated in galaxies.

Dwarf spheroidal galaxies, located at the faint and challenging end of the galaxy
luminosity function, have essentially only been targeted in their  centres where
the most recent star formation is concentrated, hence probing the end of their
star formation history and chemical evolution \cite[e.g.][]{THT09}.  The first
evolutionary steps of these systems remain mostly unexplored, yet focusing on the
extremely metal-poor regime ([Fe/H] $\le-3$) of these first evolutionary steps
adds two major dimensions to chemical evolution studies.  One is related to the
initial mass function (IMF) in the early stages of galaxy formation.  Whether or
not the IMF is universal is of critical importance, and provides deep insights into
star formation processes.  How many massive (20--100~M$_{\odot}$) stars can a
dwarf system of final stellar mass $10^5$--$10^7$M$_{\odot}$ form is a puzzle.
Only the analysis of [$\alpha$/Fe] at very low metallicity, i.e. the Type II
supernovae (SNeII)-dominated regime, is discriminant.  The second dimension
addresses the mixing of the SNe ejecta in the interstellar medium (ISM).  The
binding energy of dwarf systems is low, and the turbulence induced by the
supernova explosions can generate pockets of ISM with very different levels of
chemical enrichment, which would reveal themselves as large, observable
dispersions in stellar abundance ratios.

Unfortunately,  we know of at most five [Fe/H]$<-3$ stars per galaxy for
which we have high-resolution spectroscopy, for example   in Sextans, Sculptor, Fornax,
Ursa Minor and Draco \citep{aoki2009, cohenhuang2009, tafel2010, frebel2010a,
  cohenhuang2010,simon2015}, as well as in the ultra-faint dwarfs, Bo\"{o}tes
\citep{norris2010a, feltzing2009, ishigaki2014}, Segue~I \citep{norris2010b,
  frebel2014}, Leo IV \citep{simon2010}, and Ursa Major~II \citep{frebel2010b}.
This dearth of data spans a factor of 100 in mass-to-light ratio, preventing
any global analysis of how star formation starts up and is sustained in these
systems.

This paper presents the high-resolution spectroscopic analysis of five new
metal-poor stars in the Local Group dwarf spheroidal galaxy Sculptor.  Combining
this new dataset with earlier work, there are now ten stars that have been observed at high
spectroscopic resolution, and fourteen in total accounting for the medium-resolution analysis of \cite{starken2013}, making Sculptor the first dwarf
spheroidal galaxy in which trends and dispersion in the very metal-poor regime can
start being established in some detail.  These trends reveal the nature of the
first generations of stars (e.g. mass, numbers, spatial distribution), and the
level of homogeneity of the primitive interstellar medium (e.g. size/mass of star
forming regions, nature and energy of the explosion of supernovae).

This paper is structured as follows: Section \ref{sample} introduces the selection
of our sample and its general properties. Section \ref{stelparam} presents the
determination of the stellar atmospheric parameters and the calculations of the
abundances. Section \ref{specific} describes the specific treatment required by
some of the elements. Section \ref{discussion} presents our results and
discusses their implications. We end this paper with a short summary of our findings
in Section \ref{conclusion}.

\section{Sample}
\label{sample}

\subsection{Selection, observations, and data reduction}

\begin{figure}
\centering
\includegraphics[width=\hsize]{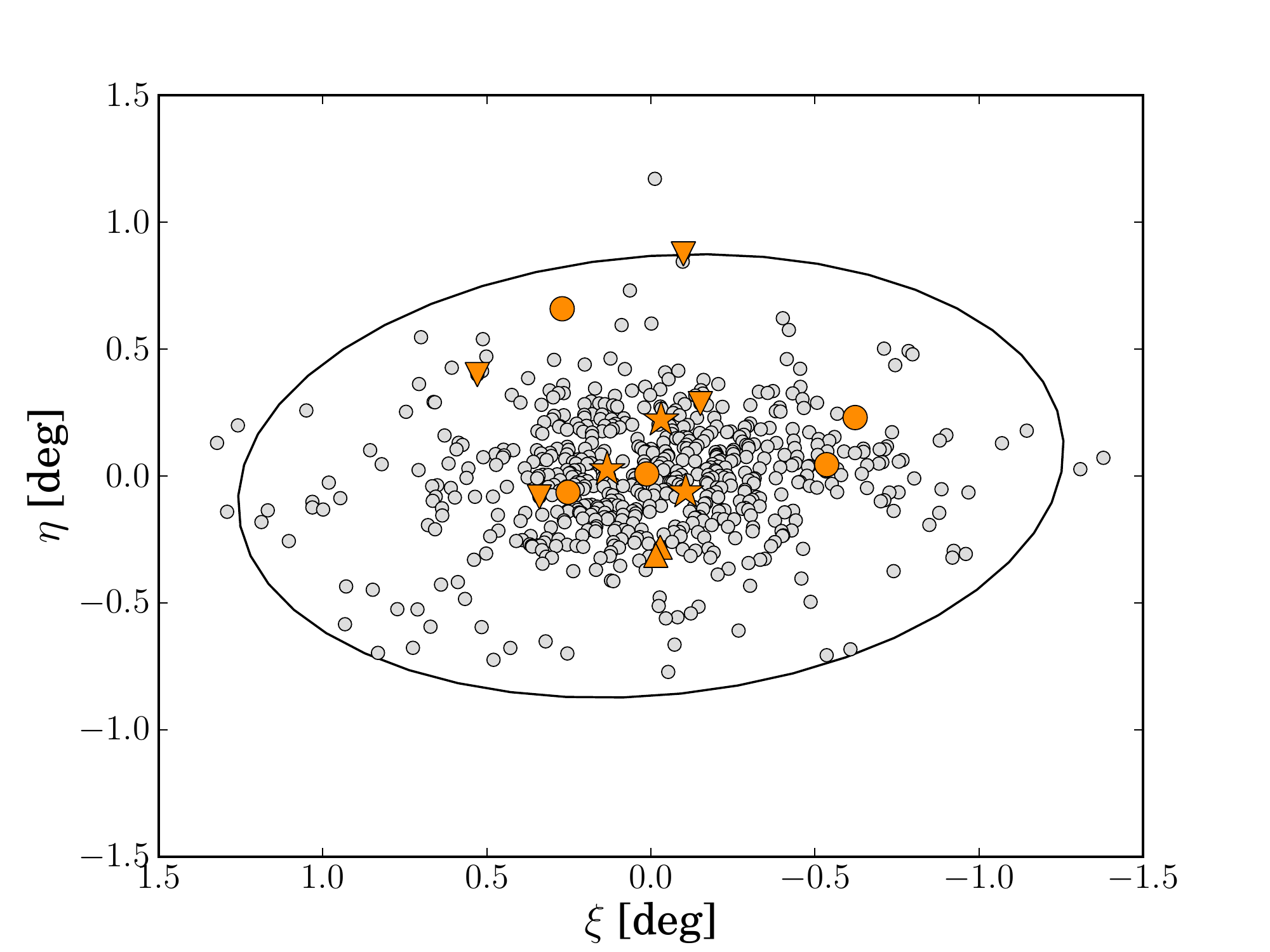}
\caption{Spatial distribution of the EMPS analyzed in this paper. The grey
  circles show the Sculptor member stars from the CaT analysis from \citet{BHT08}.
  The orange circles indicate our new sample of stars with chemical abundances
  derived from high-resolution spectroscopy. The star symbols identify the
  EMPS of \citet{frebel2010a}  and \citet{simon2015}, while the upright triangles points to the sample of
  \citet{tafel2010} and inverted triangles the EMPS of \citet{starken2013}, which
  have not been reanalyzed at high resolution.}
\label{spatial}
\end{figure}

\begin{figure}
\centering
\includegraphics[width=\hsize]{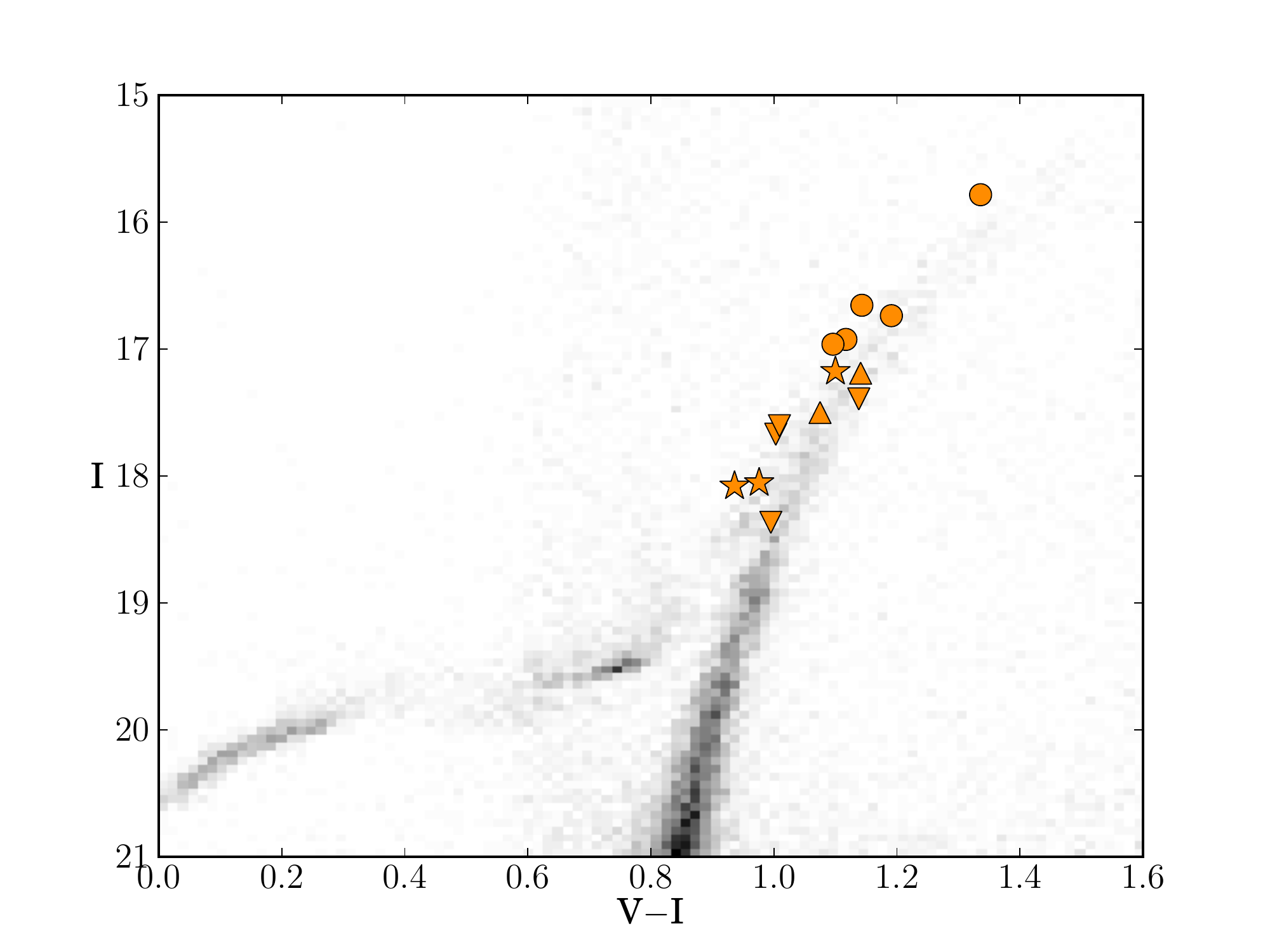}
\caption{ $V-I$ versus $I$ colour-magnitude diagram of Sculptor on which we superimpose the
  positions of the EMPS discussed in this paper. The photometry is taken from
  \citet{BTS11}. The orange circles indicate our new sample of stars with chemical abundances
  derived from high-resolution spectroscopy. The star symbols identify the
  EMPS of \citet{frebel2010a}  and \citet{simon2015}, while the upright triangles points to the sample of
  \citet{tafel2010} and inverted triangles the EMPS of \citet{starken2013}, which
  have not been reanalyzed at high resolution.}
\label{cmd}
\end{figure}

Our sample was drawn from the medium-resolution CaT survey of
\citet{BHT08}. \citet{SHT10} provided the community with a new CaT calibration, which
enables stellar metallicity to be estimated down to [Fe/H] $=-4$. We selected the
probable extremely metal-poor stars (EMPS), which were bright enough to be observed
at high resolution.  Figure \ref{spatial} gives their spatial distribution and
Fig. \ref{cmd} their location on the red giant branch (RGB) of the Sculptor
dSph. Three stars, scl002\_06, scl031\_11, and scl074\_02 had been  observed previously at
medium resolution with Xshooter by \citet{starken2013}. The high-resolution
spectroscopy brings a number of new and crucial elements, such as 
Co, Al, Si, Sc, and Mn. We conduct a brief comparison between the results of the
two studies in Sect. \ref{comparison}.

The observations were conducted in service mode with the UVES spectrograph attached
to the VLT second unit, Kueyen. The slit width was set to 1\arcsec, ensuring a
resolving power $R=45000$ between $\sim$3500\AA\ and $\sim$6850\AA. The journal of the
observations, including target names, coordinates, the useable
wavelength range for each star, signal-to-noise ratios per wavelength range, and
exposure times is given in Table~\ref{table:journal_obs}.

The reduction was done with the ESO UVES pipeline (release 5.09) with optimal
extraction. The 1D spectra resulting from order merging were then visually
examined and the remaining obvious cosmic rays were removed by hand, using the IRAF splot
subroutine\footnote{IRAF is distributed by the National Optical Astronomy Observatory
  (NOAO), which is operated by the Association of Universities for Research in
  Astronomy (AURA), Inc., under cooperative agreement with the U.S. National Science
  Foundation}.

\subsection{Radial velocities and equivalent widths}
\label{howlines}
Each of the three UVES wavelength ranges was normalized using
DAOSPEC\footnote{DAOSPEC has been written by P. B. Stetson for the Dominion
  Astrophysical Observatory of the Herzberg Institute of Astrophysics, National
  Research Council, Canada.} \citep{SP08,SP10}, and subsequently 4DAO\footnote{4DAO
  is a FORTRAN code designed to launch DAOSPEC automatically for a large sample of
  spectra.}.  This code allows some spectral regions (e.g. telluric lines,
residuals of sky lines) to be masked and displays the Gaussian fit to each individual
line \citep{M13}.  For the two red spectral ranges, the order merging resulted in a
flux modulation requiring a high-order polynomial fit of the continuum. In the few
cases where large amplitude wiggles still remained, the continuum was then placed
manually and the equivalent widths (EQWs) were recalculated with Gaussian fits or
direct integration with the {\sf iraf splot} routine.

 The DAOSPEC and 4DAO codes actually fit {\sl saturated} Gaussians to the strong
 lines, not simple Gaussians. Nevertheless, they can not fit the wide,
 Lorentz-like wings of the profile of very strong lines, in particular beyond
 $200$\,m\AA. This is especially the case at very high resolution
 \citep{KC12}. Therefore, we systematically measured all strong lines manually, using
 Gaussian fits as well as direct integration. When the DAOSPEC estimates agreed with
 these manual measurements within the DAOSPEC error bars, they were kept. This was
 in the majority of the cases. Otherwise, we adopted the manual value closest to the
 DAOSPEC measurement. Direct integration was preferred when the wings of strong lines
 were too poorly fitted by a Gaussian.

In all cases we adopted the error $\delta EQW$ computed by DAOSPEC. It is given by
\citet{SP08}
$$\delta EQW=
\Delta\lambda^2\,\sqrt{\sum_i(\delta I_i)^2\left(\frac{\partial EQW}{\partial I_i}\right)^2
+\sum_i\left(\delta I_{C_i}\right)^2\left(\frac{\partial EQW}{\partial I_{C_i}}\right)^2},
$$
where $I_i$ and $\delta I_i$ are the intensity of the observed line profile at
pixel $i$ and its uncertainty, while $I_{C_i}$ and $\delta I_{C_i}$ are the intensity
and uncertainty of the corresponding continuum. The uncertainties on the intensities
are estimated from the scatter of the residuals that remain after subtraction of
the fitted line (or lines, in the case of blends). This is  a lower limit to
the real EQW error because systematic errors like the continuum placement are
not accounted for. The EQWs are provided in Table~\ref{table:lines_abund}.

The radial velocities (RV) were calculated with 4DAO in each spectral range on the
normalized spectra in which the telluric features were masked.  For a given star,
the RVs obtained from the three spectral ranges agree to within $\pm 0.7$\kms\ (Table~\ref{table:journal_obs}).  The average RV of each star coincides
with that of the Sculptor dSph galaxy ($110.6\pm0.5$\kms) within three times the
velocity dispersion, $\sigma=10.1\pm0.3$\kms\ measured by \cite{BHT08}, meaning that our
stars are highly probable members.

\section{Stellar parameters}
\label{stelparam}

\subsection{Models, codes, and ingredients}

We adopted the MARCS 1D spherical atmosphere models with standard abundances.  They
were downloaded from the MARCS web site\footnote{marcs.astro.uu.se} \citep{GEE08},
and interpolated using Thomas Masseron's interpol\_modeles code available on the same
web site. The models are all computed for [$\alpha$/Fe] $=+0.4$, which is perfectly
suited for all our stars except ET0381.  We were able to verify for this star that a model
with [$\alpha$/Fe]=+0.0 gives an FeI abundance larger by $0.03$~dex and a
microturbulence velocity larger by $0.05$~\kms.  Unfortunately, models with
[$\alpha$/Fe]=$-$0.4 are not available, but we can expect a systematic shift well
within the uncertainties based on the differences between the enhanced and solar
alpha models.

\begin{figure}[h]
\centering
\includegraphics[width=\hsize]{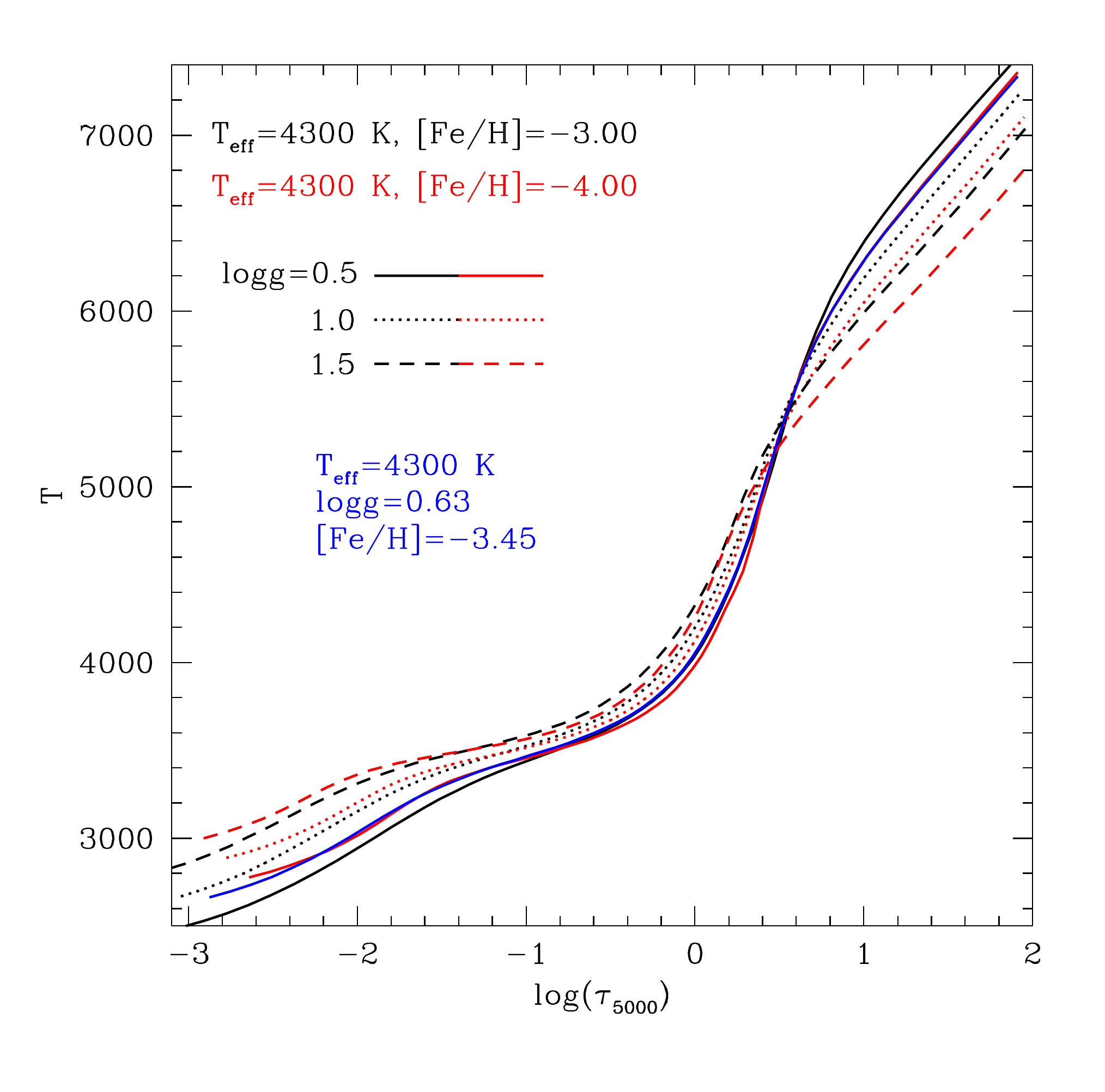}
\caption{\Teff - log($\tau$) relation for the extrapolated model of scl002\_06 (blue), compared with
grid models for [Fe/H]=$-3.0$ (black) and $-4.0$ (red), and \logg $= 0.5, 1.0, 1.5$, for \Teff=4300\,K.
All these models are interpolated in \Teff between $4250$ and $4500$\,K, but those with \logg$< 1.0$
and [Fe/H]=$-4.0$ are interpolated between eight models $(4250, 4500; 0.5, 1.0; -4.0, -3.0)$, two of
which are extrapolated $(4250, 4500; 0.5; -4.0)$ from the MARCS grid. By chance, the model
interpolated for scl002\_06 (full blue curve) almost exactly coincides with the $(4300; 0.5; -4.0)$
model (full red curve), except in the range $-0.5 <$ log($\tau_{5000}< +0.4$.}
\label{fig:extrapol}
\end{figure}

The star scl002\_06 has the lowest surface gravity in our sample and unfortunately
there is no available MARCS model at [Fe/H] $<-3.0$, \logg$<1.0$ and
\Teff$<4500$~K.  At these extremely low metallicities models suffer
numerical instabilities and it is exceedingly difficult to make them converge 
  (B. Plez, private communication).  In order to minimize systematic errors, we
therefore chose to extrapolate the model grid for [Fe/H]=$-4.0$ along the \logg
axis for \Teff$=4000$, $4250$, and $4500$~K. The extrapolation was a linear one,
applied to the logarithmic quantities ([Fe/H], \loggno, and log\Teffno) and based
on the models with \logg$=1.5$ and $1.0$. Three more models were defined in this
way, with \logg$=0.5$, [Fe/H] =$-4.0$, [$\alpha$/Fe]=$+0.4$, and \Teff$=4000$,
$4250$, and $4500$~K.  The \Teff versus $\tau_{5000}$ relation of an example
  of extrapolation is displayed in Fig. \ref{fig:extrapol}, showing a smooth
  variation from one model to the other. Extrapolations along the [Fe/H] and the
\Teff axis were also tested, but were less convincing.  Then Masseron's
interpolating code was used with the extended grid.

The abundance analysis and the spectral synthesis calculation were performed with the
turbospectrum code \citep{AP98,P12}, which assumes local thermodynamic equilibrium
(LTE), but treats continuum scattering in the source function. We used a plane
parallel transfer for the line computation to be consistent with our previous work on
EMPS \citet{tafel2010}. The same code was used to produce synthetic spectra of short
spectral intervals with various abundances and to estimate the C abundance by visual
interpolation in the G-band of the CH molecule.

The adopted solar abundances in Table~\ref{table:mean_abund} are from
\citet{anders89} and \citet{grevesse98}.  Our line list combines those of
\citet{tafel2010} and \citet{vanderswaelmen2013}, with the exception of the
  CH molecule for which we used the list published by \citet{MP14}. The turbospectrum
code was fed with information on the spectral lines taken from the VALD database
\citep{PKR95, RPK97, KPR99, KRP00}. However, we kept the $\log(gf)$ values from
the original list. The central wavelengths and oscillator strengths are given in
Table~\ref{table:lines_abund}.

\subsection{First photometric approximations to the effective temperature and surface gravity}

The first approximation of the stellar effective temperature was based on the $V$- and
$I$-band magnitudes measured in \citet{BTS11,BTH12} and based on $J$ and $K_s$
photometry taken from the VISTA commissioning data, which was also calibrated onto
the 2MASS photometric system.

We assumed Av=3.24 $\cdot$ E$_{\mathrm{B-V}}$ \citep{cardelli1989} and E$_{\mathrm{B-V}}
=0.018$ \citep{schlegel1998}.  The final photometric \Teff for each star indicated in
Table~\ref{table:photometry} corresponds to the simple average of the three-colour
temperatures derived from $V-I$, $V-J$, and $V-K$ with the calibration of
\cite{RM05}.

Because of the very small number of detectable \ion{Fe}{ii} lines, we determined \logg\ from
its relation with \Teff using the calibration for the bolometric correction of
\cite{AAM99}, 

\noindent $\logg_{\star} = \logg_{\odot} +
\log{\frac{\mathcal{M}_{\star}}{\mathcal{M}_{\odot}}} + 
4 \times \log \frac{T_{\mathrm{eff}\star}}{T_{\mathrm{eff}\odot}} + 0.4 \times (M_{\mathrm{Bol}\star} - M_{\mathrm{Bol}\odot})$,
\label{eq:logg}
where $\log g_\odot=4.44$, $T_\mathrm{eff\odot}=5790$~K, M$_{bol\odot}=4.75$, and
${\cal M}=0.8~{\cal M}_\odot$. The distance of \citet{PGS08}  $d=85.9$~kpc,
was adopted to calculate M$_{bol}$.

\subsection{Final stellar parameters}

The convergence to our final effective temperatures and microturbulence velocities
($v_\mathrm{t}$) presented in Table \ref{table:atmo} was achieved iteratively as a
trade off between minimizing the trends of metallicity derived from the \ion{Fe}{i}
lines with excitation potentials ($\chi_\mathrm{exc}$) and reduced equivalent widths,
log(EQW/$\lambda$), and minimizing the difference between \ion{Fe}{ii} and
\ion{Fe}{i} abundances on the other hand.  Starting from the initial photometric
parameters we adjusted \Teff\ and \vmic\ by minimizing the slopes of the diagnostic
plots, allowing the slope to deviate from zero by no more than 2 $\sigma$, the
uncertainties on the slopes. New values of \logg\ were then computed from the equation above.

Following \citet{tafel2010}, we only considered the Fe lines with $\chi_\mathrm{exc}>
1.4$~eV in order to avoid 3D effects as much as possible.  We also discarded 
lines that were too weak (defined as EQW $< 20$~m\AA\ in the red and EQW $< 30-40$~m\AA\ in the blue)
because their measurements were noisier than the rest. We rejected all iron lines with EQW
$> 200$~\AA\, to minimize biases on the stellar parameters because such lines differ
too much from a Gaussian shape. However, we did keep a few such strong lines for
interesting elements such as Mg, but checked the corresponding abundances through
synthetic spectra, because as discussed in Sect. \ref{howlines}, their non-Gaussian
shape may induce a bias in the EQW determination.  Finally we used the {\sl
  predicted} EQWs in the \ion{Fe}{i} abundance versus log(EQW/$\lambda$) diagram,
following \cite{M84}. Although this does not change the results in a very significant
way, it does reduce $v_\mathrm{t}$ by $0.1-0.3$\kms and increase [Fe/H] by a few
hundredths of a dex in a systematic way,  compared to using the observed equivalent
widths.

The final stellar parameters are given in Table~\ref{table:atmo}.  The typical
errors are $\sim 100$~K on \Teffno, $\sim 0.1$~dex on $\log g$, assuming a
$\pm0.1$~\msol error on $\cal{M} _{\star}$ and a $0.2$~mag error on
M$_\mathrm{bol}$, and about $0.2$~\kms on $v_\mathrm{t}$.

\subsection{Hyperfine structure}
The hyperfine structure (HFS) broadens the line profile and tends to increase its EQW for a
given abundance because it tends to de-saturate the line. Therefore, the abundances of
elements with a significant HFS broadening, such as Sc, Mn, Co, and Ba are biased when
they are determined from the line EQWs, if HFS is neglected.

We determined the HFS correction to the abundance related to each line of the
elements concerned, by running Chris Sneden's MOOG
code\footnote{http://www.as.utexas.edu/~chris/moog.html} with the \verb+blend+
driver on a line list including the HFS components, as in \citet{NCJ12}. The HFS
components with their oscillator strengths were taken from \citet{PMW00} for Sc
and Mn, and from the Kurucz web
site\footnote{http://kurucz.harvard.edu/linelists.html} for Co and Ba.
The HFS correction is small for weak lines (e.g. for the \ion{Ba}{ii} subordinate
lines), but may reach $-0.5$~dex for \ion{Mn}{i}.

\subsection{Final abundances}

The final abundances are calculated as the weighted mean of the abundances
obtained from the individual lines, where the weights are the inverse variances of
the single line abundances.  These variances were propagated by turbospectrum from
the estimated errors on the corresponding equivalent widths. The average
abundances based on EQWs and  the C abundances are given in
Table~\ref{table:mean_abund}.

The upper limits to the Eu, Ba, Y, or Zn abundances are also provided. They
  are based on visual inspection of the observed spectrum, on which seven synthetic
  spectra were overplotted, with abundance varying by steps of 0.1~dex. The upper
  limit adopted corresponds to a synthetic line lying at the level of about
  $-1\,\sigma$, where $\sigma$ is the rms scatter of the continuum.

\subsubsection{Errors}

In order to make the errors on EQW more realistic, we added quadratically a 5\%
error to the EQW error estimated by DAOSPEC, so that no EQW has an error
smaller than 5\%. The errors estimated in this way were found to be generally larger
than those obtained from the \citet{C88} formula revised by \citet{BIT08}. They
are given in Table~\ref{table:lines_abund}.

The $\sigma_\mathrm{EQW}$ errors listed in Table~\ref{table:mean_abund} are defined
in the same way as in \citet{tafel2010} and \citet{starken2013}. The average
abundance error due to the EQW error alone for one average line is
$\sigma_\mathrm{EQW}$.  Here, it is computed as
$\sigma_\mathrm{EQW}=\sqrt{\frac{N}{\sum_i 1/\sigma_i^2}}$, where N is the number of
lines.  The $\sigma_\mathrm{X}$ errors correspond to the rms scatter of the
individual line abundances, divided by the square root of the number of lines, $N$:
$\sigma_\mathrm{X}=\sqrt{\frac{\sum_i(\epsilon_i-\overline{\epsilon})^2}{N(N-1)}}$.
The final error on the average abundances is defined as
$\sigma_\mathrm{fin}=max\left(\sigma_\mathrm{EQW},\sigma_\mathrm{X},
\sigma_\mathrm{Fe}\right)$. As a consequence, no element $X$ can have
$\sigma_\mathrm{X} < \sigma_\mathrm{Fe}$; this is particularly important for species
with a very small number of lines.

The errors provided in Table~\ref{table:lines_abund} do not include the propagation
of the errors on the stellar parameters, especially \Teffno. For this purpose, we
give the effect of a $100$~K \Teff increase on mean abundances in
Table~\ref{table:Tp_del_abun}. Since \logg\ and $v_\mathrm{t}$ change according
to \Teffno, we do not consider them to be independent. Therefore, the impact of a variation in \Teff on
the abundances is given after \logg\ was adapted to the new \Teff value and
$v_\mathrm{t}$ was optimized on the basis of the \ion{Fe}{i} abundance
versus log(EQW/$\lambda$) diagram.

\section{Comments on specific abundances}
\label{specific}

Table~\ref {table:lines_abund} presents the measurement of the equivalent widths
of the lines that have been considered in the analysis and their corresponding
elemental abundances.  In the next few subsections, we address a few distinctive
points. Elements and stars that did not require any specific treatment or that
gave consistent results for all lines and are not prone to any possible biases
such as the non-local thermodynamic equilibrium (NLTE) effect do not appear in
this section and we refer the reader to Table~\ref {table:lines_abund} and Section
5.

\subsection{Carbon}
\begin{figure}
\centering
\includegraphics[width=6cm]{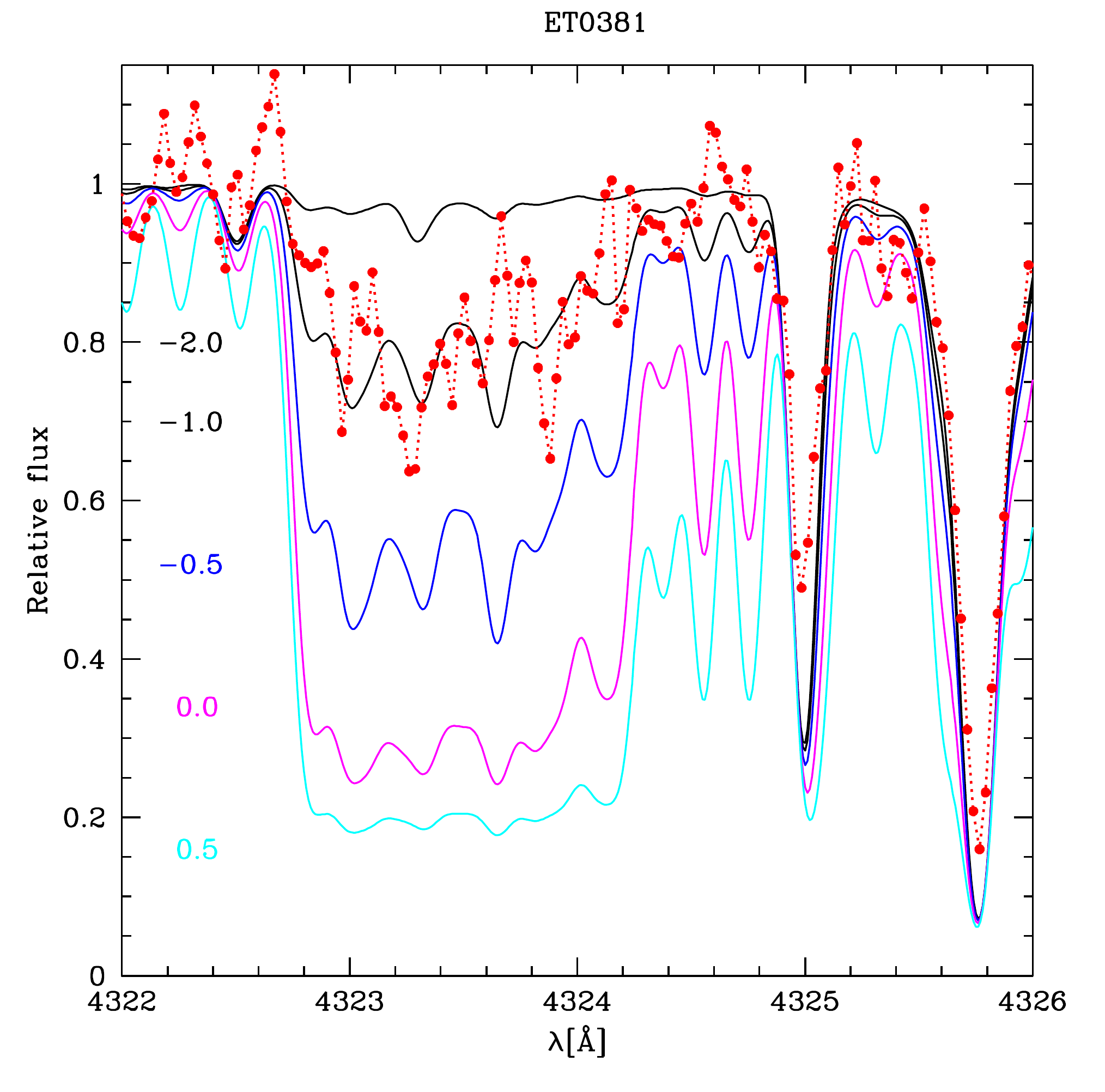}
\caption{Observed spectrum of ET0381 in a short spectral region of the molecular
CH band. Synthetic spectra with [C/Fe]$=-2.0$ to $+0.5$ are shown as well. The numbers
indicated on the left are the corresponding [C/Fe] values.}
\label{fig:CsFe_ET0381}
\end{figure}

The C abundance is based on the intensity of the CH molecular feature at
$4323-4324$~\AA\ in the G band. To compute a synthesis in this region, the oxygen and
nitrogen abundances have to be known or assumed, since part of the carbon is locked
in the CO and CN molecules. As do \citet{tafel2010} and \citet{starken2013}, we
assume that the [O/Fe] ratio is the same as [Mg/Fe] and that [N/Fe] is solar, because
we are unable to measure the oxygen and nitrogen abundance.  We also adopt
$^{12}\mathrm{C}/^{13}\mathrm{C}=6$, an appropriate value to the tip of the
RGB \citep{SCH06}. Figure~\ref{fig:CsFe_ET0381} shows a comparison between five synthetic spectra with
increasing C abundance and the observed spectrum of the star ET0381.

\subsection{$\alpha$ elements}

\begin{itemize}
\item {{\it Magnesium}} The Mg abundance is based on $4$ or $6$ lines distributed
  from the violet to the yellow part of the spectrum. Four of them are strong,
  with EQW $>100$~m\AA, while the other two are much weaker, nevertheless they all
  provide consistent abundances. We avoided the \ion{Mg}{i}
$\lambda 3829$\AA\ line because it is strongly blended.

\item {{\it Titanium}} The \ion{Ti}{i} abundances rely on $4$ to $11$ faint lines
  giving consistent values, while the \ion{Ti}{ii} abundances are based on $17$ to
  $28$ lines, many of them stronger, and giving more scattered results. We adopt the
  \ion{Ti}{ii} abundances as representative of titanium because they are less
  sensitive to NLTE effects.
\end{itemize}

\subsection{Iron peak elements}

\begin{itemize}

\item {{\it Manganese}} The abundance is based on $3$ to $5$ lines.  They are
  corrected for HFS. The strong, blue resonance lines at $\lambda4030$\AA, $4033$\AA,
  and $4034$\AA\ suffer from NLTE effects and from some blends, but they are the only
  ones available in the most metal poor stars. To minimize the effects of blends,
  we did not consider $\lambda4033,\,4034$\AA\ for ET0381.  In the three stars where
  blue resonance lines ($\lambda 4030$\AA, $4033$\AA, and $4034$\AA) and subordinate ones
  ($\lambda 4041$\AA, $4823$\AA) could be measured, the abundances resulting from the
  former are $0.2-0.3$~dex lower than those resulting from the latter. This is
  qualitatively consistent with, but less pronounced than, the $0.5$~dex
  difference seen by Venn et al. (2012) for their own stars.

\item {{\it Nickel}}
The abundances are based on $2$ to $3$ lines, avoiding the blended $\lambda4231$\AA\ line.

\item {{\it Zinc}}
The only usable line is the very faint \ion{Zn}{i} at $\lambda 4810$\AA~, and
only an upper limit to the abundance could be determined.

\end{itemize}

\subsection{Neutron capture elements}

\begin{itemize}

\item {{\it Strontium}}
The Sr abundance is based on two resonance lines, \ion{Sr}{ii} $\lambda 4077$\AA\ and
$\lambda 4215$\AA. The latter  is blended primarily with the \ion{Fe}{i}
$\lambda4215.423$\AA\ line, and also, to a much lesser extent, by faint molecular
CN lines. Because that iron line is relatively strong for ET0381, according to its  spectral
synthesis, we did not consider  \ion{Sr}{ii} $\lambda 4215$\AA~ in that star.

\item {{\it Europium}}
Only one line, \ion{Eu}{ii} $\lambda$4129\AA\, can be used in our spectral range, though
in all cases it was too faint to be detected. Thus we are only able to give upper
limits to the Eu abundances. 

\item {{\it Barium}} 
Barium is represented by at least five isotopes with significant contributions
from the odd atomic mass ($A$) nuclei. Indeed the ratio between the even-$A$ and
the odd isotopes n(\iso{134}+\iso{136}+\iso{138}) : n(\iso{135}+\iso{137}) is
82:18, or in other words the fraction of odd-$A$ isotopes, $f_\mathrm{odd}$, is
0.18 in the solar system \citep{Lodders2009}. While for the odd-$A$ isotopes the
nucleon-electron spin interactions lead to hyper-fine splitting of the energy
levels, the even isotopes are unaffected by HFS. The level splitting of
\ion{Ba}{ii} reaches its maximum at ground state, and the components of the
resonance line 4934\,{\AA} are separated by
78\,m\AA\ \citep{Brix1952,1983ZPhyA.311...41B,1982PhRvA..25.1476B,1980ZPhyA.295..311S},
and the Ba abundance derived from this line depends on the isotope mixture adopted
in the calculations. Consequently, the barium abundances of ET0381 and
scl\_03\_059 were determined from the \ion{Ba}{ii} subordinate lines \ion{Ba}{ii}
5853, 6141, and 6497\,{\AA}, which are hardly affected by HFS
(Table~\ref{table:lines_abund}), while the \ion{Ba}{ii} 4934\,{\AA} line was used
to determine the Ba even-to-odd isotope abundance ratios, by requiring that all
lines return the same abundance. The \ion{Ba}{ii} 4934 and 6141\,{\AA} lines were
corrected for their blend with \ion{Fe}{i} lines by synthesis. \\

It is commonly accepted that the elements heavier than Ba have a pure $r$-process
origin in the metal-poor MW halo stars \citep{truran1981}.  The $r$-process model of
\citet{Kratz2007} predicts $f_\mathrm{odd}$ = 0.438 in the classical waiting-point
(WP) approximation. Another estimate of $f_\mathrm{odd}$ is provided by the analysis
of the contribution of the r-process to the solar system (SS) Ba abundance and
subtraction of the s-process contribution from the solar total abundance, with a
range of possible fractions, $f_\mathrm{odd}$ = 0.46 \citep{Travaglio1999}, 0.522
\citep{sneden1996}, 0.601 \citep{Bisterzo2014}, and 0.72 \citep{McWilliam1998}.

We first considered $f_\mathrm{odd}$ = 0.46 and 0.72 to derive the \ion{Ba}{ii}
abundance from the $\lambda 4934$\AA\ line. In both cases and for both ET0381 and
scl\_03\_059, we found significant discrepancies between the abundances derived
from the subordinate lines and from the resonance line
(Table~\ref{Tab:hfs}). Further investigation shows that these differences are
removed for $f_\mathrm{odd} = 0.18$, i.e. the solar Ba isotope mixture
\citep{Lodders2009} or even $f_\mathrm{odd} = 0.11$, which is predicted by \cite{Bisterzo2014}
for pure s-process production.

\begin{figure}
\centering
\includegraphics[width=\hsize]{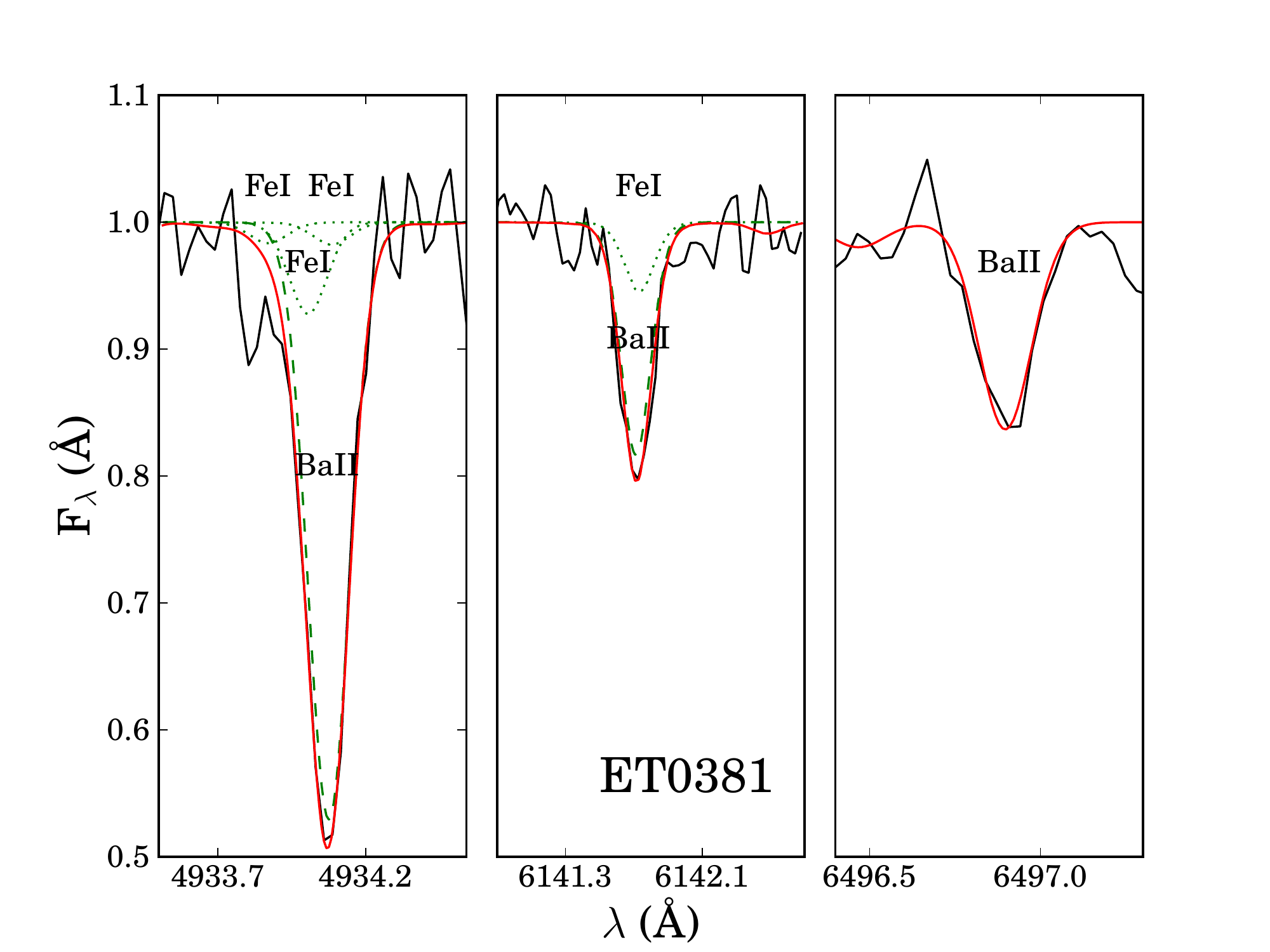}
\includegraphics[width=\hsize]{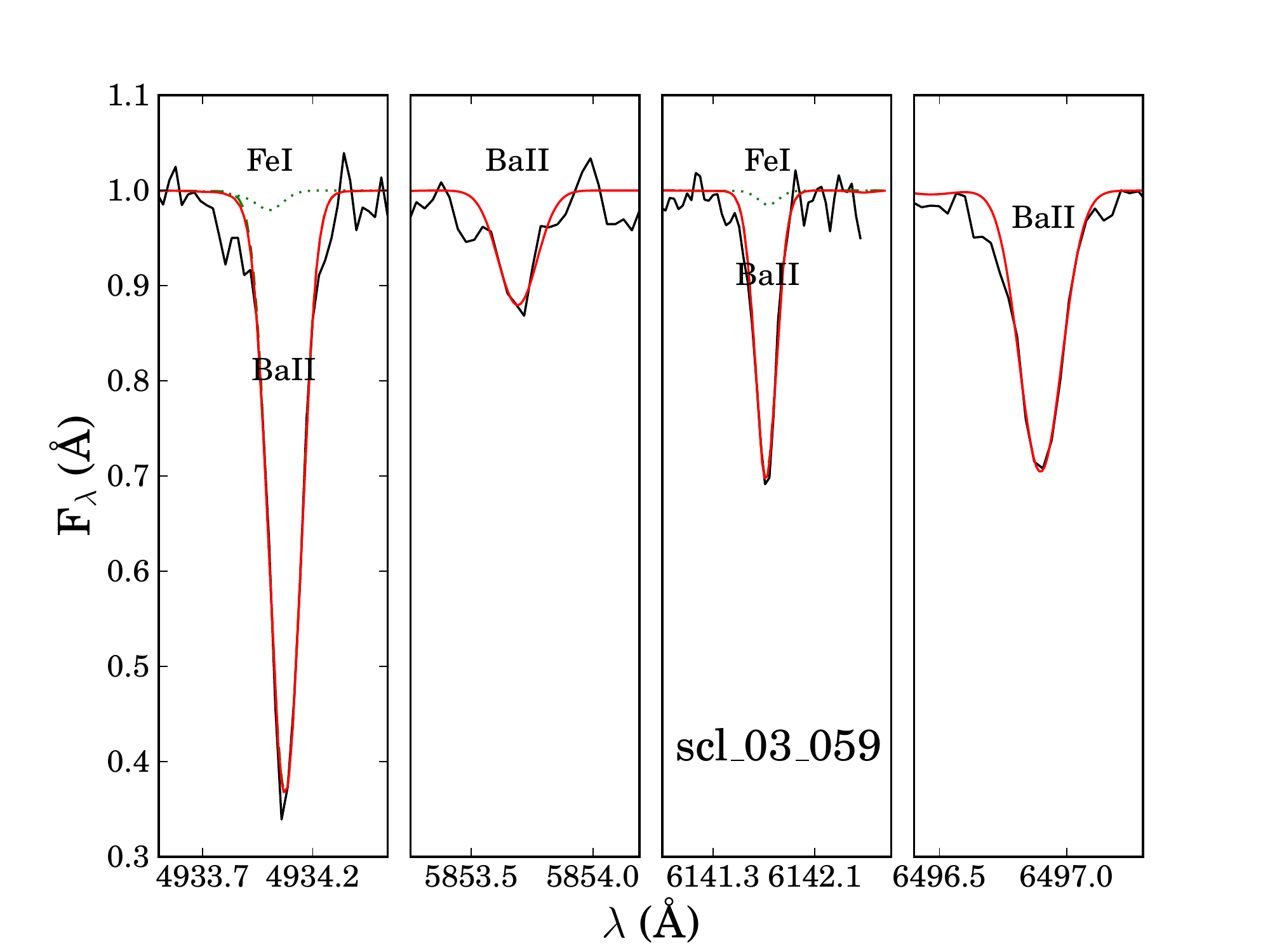}
\caption{Synthesis of the Ba lines for ET0381 (upper panel) and scl\_03\_059 (lower
  panel). A value of $f_\mathrm{odd}$ = 0.18 is taken for the \ion{Ba}{ii} 4934\,\AA. The
  observed spectra are shown in black. The red continuous, green dashed, and green
  dotted curves correspond to the theoretical spectra of the full blend, pure \ion{Ba}{ii}
  lines, and \ion{Fe}{i} blending lines, respectively.}
\label{fig:Basynth}
\end{figure}

We performed a series of tests to estimate how much of the differences between the
subordinate and the \ion{Ba}{ii} 4934\,{\AA} lines found for the $r$-process Ba
isotope mixtures could simply arise from uncertainties in our analysis:

{\it i)} We tested the exact same methodology on HE1219-0312, a well-known r-process
enhanced star with [Eu/Fe]$=+1.4$ \citep{hayek2009}.  Using the spectrum reduced by
Norbert Christlieb, we find a mean abundance of the \ion{Ba}{ii} 5853\,{\AA},
6141\,{\AA}, and 6496\,{\AA}, lines to be $\eps{}$=$-0.06 \pm 0.03$ in LTE
calculation, and $-0.16 \pm 0.11$ in NLTE. The $\lambda$4934\,{\AA} line leads to an
NLTE abundance of $-0.09$ for $f_\mathrm{odd} = 0.46$ and 0.21 for $f_\mathrm{odd} =
0.18$. This rules out a low $f_\mathrm{odd}$. Our analysis is also in agreement with
the results of \citet{hayek2009}, who determined log$\epsilon$(Ba) $=-0.14$.

{\it ii)} Since the atmosphere of very metal-poor stars can be subject to deviation from LTE,
we performed  NLTE calculations, following the procedure
of \citet{Mashonkina1999}.  It turns out that the departure from LTE is very small
for the \ion{Ba}{ii} 4934\,{\AA} line, with a NLTE abundance correction of $\Delta_{\rm
  NLTE} = \eps{NLTE} - \eps{LTE}$ = 0.00 for ET0381 and $-0.01$~dex for
scl\_03\_059 (Table~\ref{Tab:hfs}). As to the subordinate lines, the NLTE
calculation leads to higher abundances by 0.05~dex to 0.1~dex, resulting in even
larger discrepancies with the \ion{Ba}{ii} 4934\,{\AA} line.

{\it iii)} Another source of uncertainty arises from the atmospheric
parameters. The consequence of a variation of each of \Teffno, log~g, or
$v_\mathrm{t}$ has been estimated (Table~\ref{Tab:hfs}).  The ``total'' quantities
correspond to the total impact of varying stellar parameters, computed as the
quadratic sum of the three sources of uncertainties. For ET0381, these
uncertainties produce very similar abundance shifts for the \ion{Ba}{ii}
4934\,{\AA} and the subordinate lines, and, therefore cannot explain the
difference between them. For scl\_03\_059, the abundance derived from the
\ion{Ba}{ii} 4934\,{\AA} line is slightly more affected than the subordinate
lines, however, not large enough yet to explain the observed discrepancy.

{\it iv)} We did not use any 3D hydrodynamical model atmospheres to derive the
abundances of barium. This element is only detected in the \ion{Ba}{ii}
majority species, and the detected lines arise either from the ground or
low-excitation levels. \citet{2013A&A...559A.102D} predicted that the (3D-1D)
abundance corrections for metal-poor giant stars are small for the \ion{Ba}{ii}
E$_\mathrm{exc}$ = 0 and 2\,eV lines: (3D$-$1D) = $-0.05$~dex and 0.0~dex,
respectively, in the 5020/2.5/$-2$ model and (3D$-$1D) = $-0.10$~dex and $-0.02$~dex
in the 5020/2.5/$-3$ model. The \ion{Ba}{ii} 5853, 6141, 6496\,{\AA} lines arise from
the E$_\mathrm{exc}$= 0.6-0.7 eV level. Interpolating between E$_\mathrm{exc}$ = 0
and 2\,eV in \citet{2013A&A...559A.102D}, one sees that the difference between the
\ion{Ba}{ii} 4934\,{\AA} and the \ion{Ba}{ii} subordinate lines would not be reduced
since the (3D-1D) abundance correction is more negative for the \ion{Ba}{ii}
4934\,{\AA} than for the other lines.

It turns out that none of the above plausible sources of errors is sufficient
to explain the difference between the subordinate and the \ion{Ba}{ii} 4934\,{\AA} lines.
 Figure \ref{fig:Basynth} displays the result of our NLTE synthesis assuming
  $f_\mathrm{odd} = 0.18$ for the Ba~II 4934\,\AA\ line. All Ba lines, which have very
  different strengths, are well reproduced 
  whether they are blended with \ion{Fe}{i} lines or not. In conclusion, one can
assert that, for ET0381 and scl\_03\_059, consistency between the different lines of
\ion{Ba}{ii} can only be achieved by applying a low fraction of the odd-$A$ Ba
isotopes, $f_\mathrm{odd}$ between 0.18 and 0.11.  At first glance this result would
imply a dominant contribution of the $s$-process to barium.  The implications and
alternative scenarios are discussed further in Section \ref{neutron}.

\end{itemize}

\section{Discussion}
\label{discussion}

\subsection{Comparison samples: References, symbols, and colour codes in figures }
\label{symbols}

In all figures, the grey  points show the location of RGB stars in the Milky Way
halo \citep{honda2004,cayrel2004, spite2005,aoki2005, cohen2013, cohen2006,
  cohen2004, spite2006,aoki2007, lai2008, yong2013, ishigaki2013}.

The Sculptor stars are shown in orange. We distinguish the new sample presented
in this paper by large circles with error bars. These error bars add in quadrature
the random and systematic uncertainties listed in Table~\ref{table:lines_abund} and
Table~\ref{table:Tp_del_abun}. The sample of \cite{tafel2010} is shown by upright
triangles, while the \cite{starken2013} stars, which were not re$-$observed at high
resolution, are shown by inverted triangles. The sample of \cite{simon2015}
including the \cite{frebel2010a} star are indicated by a star. For part of the
elements smaller orange circles, at slightly higher metallicities than the stars
discussed here, are from \cite{THT09} and Hill et al. (in prep).

Figures \ref{SrBaBaH} and \ref{CL} show the dataset of \cite{tafel2010} with
green triangles for Sextans and blue triangles for Fornax. The Ursa Minor
population is shown by purple stars \citep{cohenhuang2010}, while the two Carina 
stars from \cite{venn2012} are seen in pink circles. We indicate in Fig. \ref{CL}
the Draco stars with cyan circles \citep{cohenhuang2009, shetrone2013,
  kirby2015}.

The ultra$-$faint dwarf spheroidal galaxies are displayed with black symbols.  Ursa
Major~II from \cite{frebel2010b} is identified with upright triangles, Coma Berenices
from \cite{frebel2010b} with stars, Leo IV from \cite{simon2010} with squares,
Hercules from \cite{koch2008b, aden2011} with pointing down triangles, Segue I from
\cite{norris2010b, frebel2014} with right-pointing triangles, Bo\"{o}tes from
\cite{norris2010a,norris2010b, lai2011, feltzing2009, ishigaki2014} with diamonds.

\subsection{Comparison of medium- and high-resolution analyses}
\label{comparison}

\begin{figure}
\centering
\includegraphics[width=\hsize]{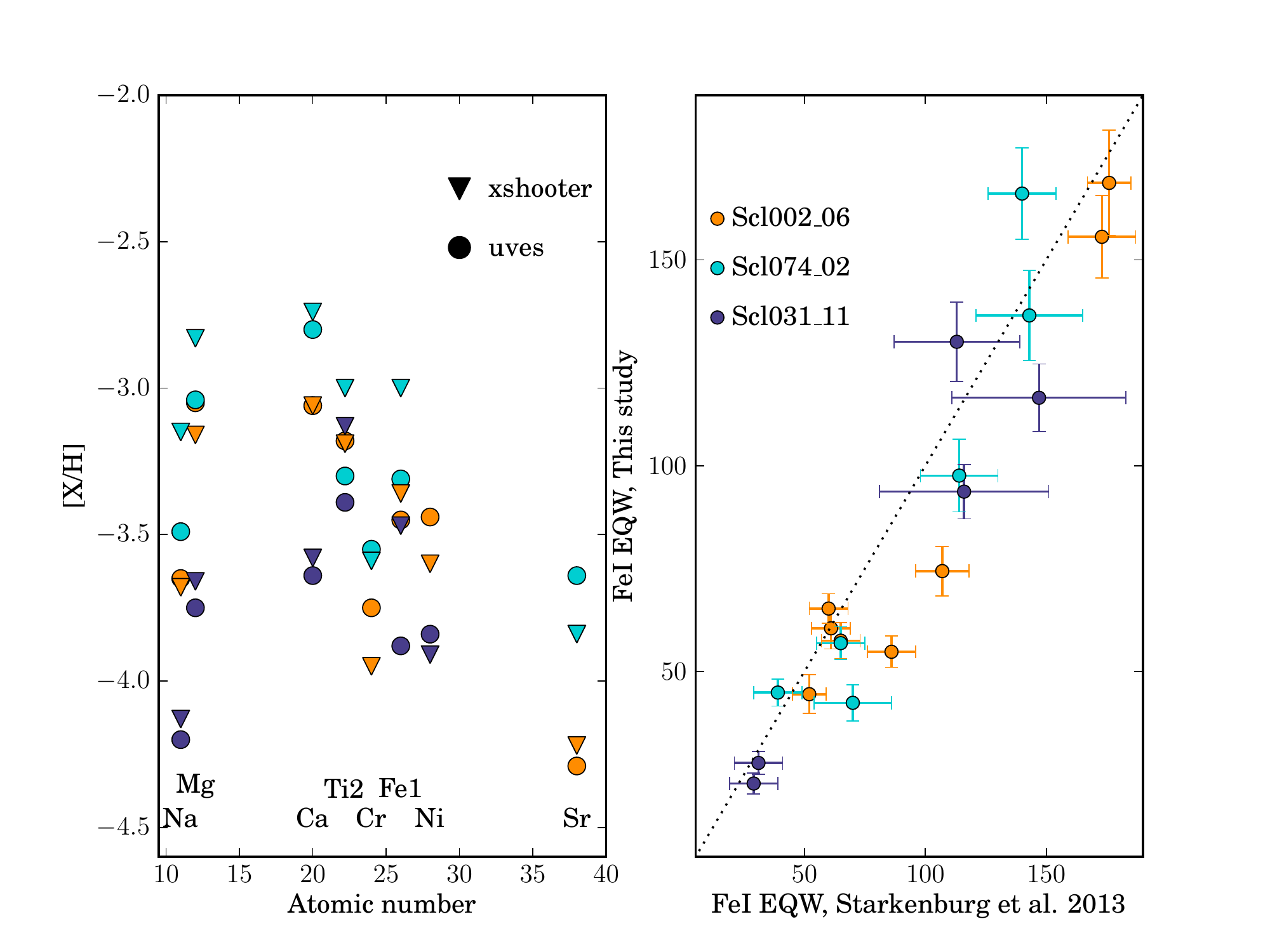}
\caption{Comparison between \citet{starken2013} and the present analysis,
   obtained from medium-resolution (Xshooter) and high-resolution (UVES)  spectra, respectively. For the
  three stars in common, the left panel compares the elemental abundances, whereas
  the right panel displays the equivalent widths of the \ion{Fe}{i} lines that were used in
  both analyses.}
\label{COMP_UVES_XHS}
\end{figure}

The left panel of Fig. \ref{COMP_UVES_XHS} compares the abundances obtained at
medium resolution by \citet{starken2013} and those of our high-resolution analysis
for the chemical elements in common.

Overall the agreement is very good, it is even excellent for scl002\_06. The
largest difference in [Fe/H] is found for scl031\_11 with $\delta$[\ion{Fe}{i}/H]
= 0.41, the present analysis having a lower metallicity.  This is a 2.5$\sigma$
shift considering the error quoted in \citet{starken2013} in particular due to the
uncertainties on the atmospheric parameters (0.16 dex total). We have better
constraints with 38 \ion{Fe}{i} lines instead of 23, and were able to get a spectroscopic
temperature  that is 100K lower than the photometric initial value. This could partly
explain the differences between the two studies. The
difference in iron abundance for scl074\_02 is less straightforward to explain
given that \Teff and \logg are identical within 5K and 0.02, respectively.  This
led us to compare the equivalent widths of the lines of \ion{Fe}{i}, which are
common to both studies. The right panel of Fig. \ref{COMP_UVES_XHS} reveals that
there is a small but systematic shift towards higher EQWs in \citet{starken2013},
for lines with EQW $>50$m\AA. These lines are located in the blue noisiest part of
the spectra.  In the present work, we find scl074\_02 and scl002\_06 at very
similar metallicities, while Starkenburg et al. determine a ~0.4 dex higher
metallicity for scl074\_02 than for scl002\_06. The Xshooter spectrum of
scl002\_06 had a mean signal-to-noise ratio of 70 against 48 for scl074\_02 in the
range 4000\AA\ to 7000\AA\ where the \ion{Fe}{i} lines are compared, leading to
slightly overestimated EQWs - as far as we can tell from the lines we have in
common - probably responsible for the higher [Fe/H] in \citet{starken2013}.

This comparison confirms the ability of medium resolution to get good accurate
abundances when it is conducted over a larger wavelength range. Potential biases
can be removed by reaching high signal-to-noise ratios.

\subsection{Carbon}

\begin{figure}
\centering
\includegraphics[width=\hsize]{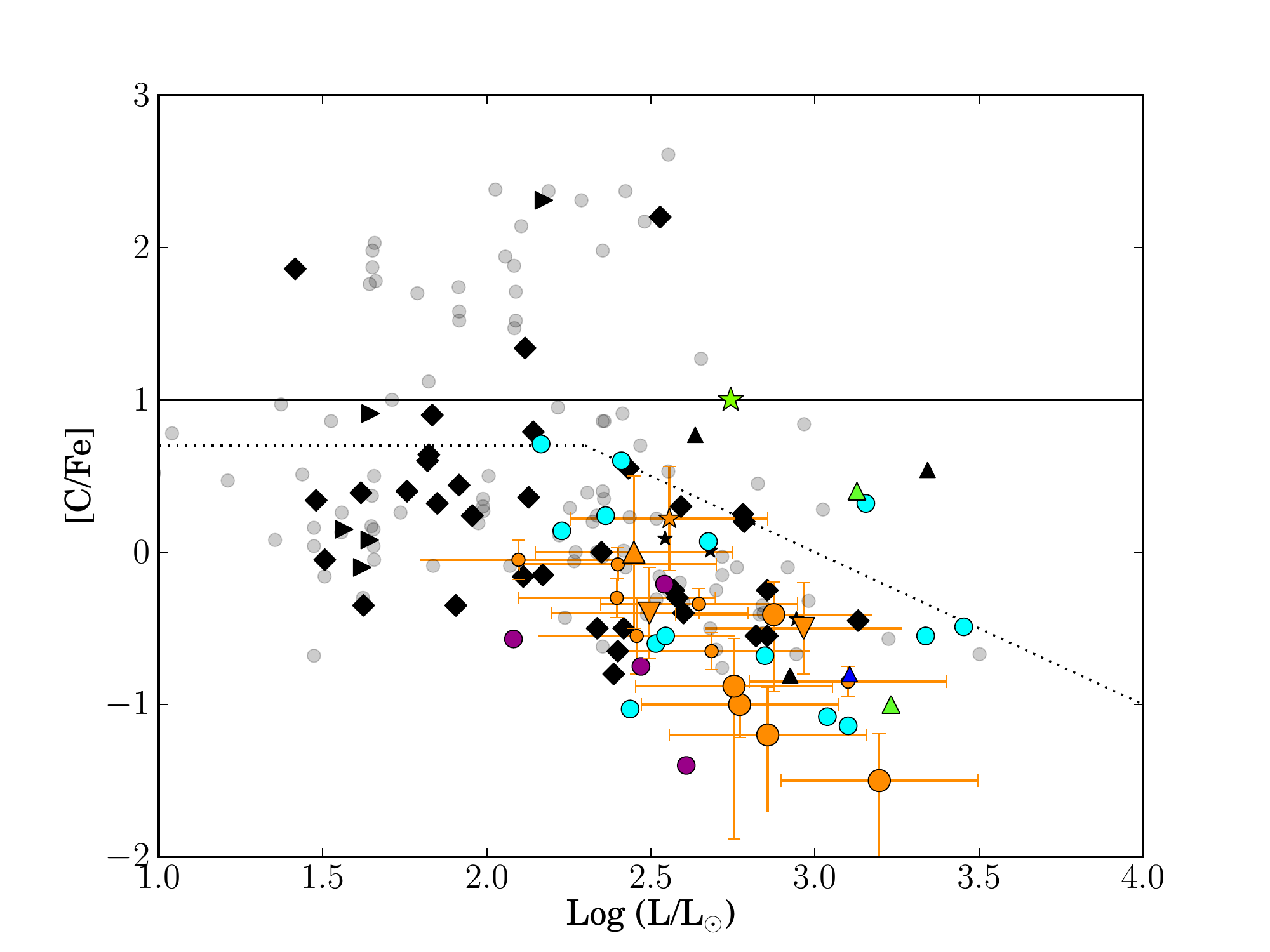}
\caption{Relation between [C/Fe] and the luminosity of the very metal-poor RGB
  stars selected as \Teff $<$5300 \AA\ and [Fe/H]$<-2.5$. The dotted line
  corresponds to the limits of \cite{aoki2007} distinguishing between carbon-rich
  and normal population. The green colour identifies the EMPS in Sextans with a
  stellar symbol for the analysis of \citet{honda2011}, and with upright triangles
  for the \citet{tafel2010} sample. The blue triangle identifies Fornax
  \citep{tafel2010}. The cyan circles show the observations in Draco by
  \citet{shetrone2013} and \citet{kirby2015}, while the point above the dotted
  line is the carbon-rich RGB found by \citet{cohenhuang2009} in this galaxy. The
  observations of \citet{kirby2015} in Ursa Minor are shown in purple. The
  references of the other comparisons samples are given in
  Sect.~\ref{symbols}. The Milky Way halo stars are shown in grey. The Sculptor stars are
  seen in orange. The errors on their luminosities are calculated for a 100K
  uncertainty in \Teff. The ultra$-$faint dwarf spheroidal galaxies are indicated
  with black symbols.  Ursa Major~II is identified with upright triangles, Coma
  Berenices with stars, Leo IV with squares, Hercules with pointing down
  triangles, Segue I with pointing right triangles, and Bo\"{o}tes with diamonds.}.
\label{CL}
\end{figure}

Figure~\ref{CL} shows how the carbon-to-iron ratio of the EMPS in Sculptor varies
with stellar luminosity. The comparison sample  is built
from stars with \Teff\ $\leq$ 5300K and [Fe/H] $<-2.5$.

It has been predicted  that the onset position along the RGB of an extra mixing between the
bottom of the stellar convective envelope and the outermost layers of the
advancing hydrogen-shell is located at
$\log{\frac{\mathcal{L}_{\star}}{\mathcal{L}_{\odot}}}\sim2.3$ for a metal-poor
0.8M$_{\odot}$ star \citep[][]{charbonnel1994, gratton2000,spite2005, eggleton08}.
Our sample of EMPS in Sculptor falls above this limit and therefore is expected to
have converted C into N by the CNO cycle. This is indeed the most likely  origin of
their low [C/Fe] ratios seen in Fig. \ref{CL}.

To date only one  carbon-enhanced metal-poor star  ([Fe/H] $\le
-2.$) with no enhancement of the main r- or s -process elements (CEMP-no star) has been
identified in Sculptor \citep{skuladottir2015}. The existence of this star would
in principle suggest that one should find many more remnants of the initial
enrichment in carbon at lower metallicity, but none has been found so far at [Fe/H]
$\le -2.5$. \cite{skuladottir2015} notice that the fraction of CEMP-no stars in
Sculptor starts deviating from the expected number derived from the statistics in
the Milky Way halo below [Fe/H] $\leq -3$.  With a $\sim$30\% fraction of CEMP
stars among the Milky Way halo EMPS \citep[see also ][]{yong2013,lee2013a,
  placco2014}, they argue that we should have observed three CEMP stars in
Sculptor in this metallicity range. Among the classical dwarfs, only in Sextans
and Draco have such stars been detected. One star in Sextans is a CEMP-s star with
[C/Fe]=1 \citep{honda2011}. The other ones are CEMP-no stars; they do not have a particularly high
value of [C/Fe] in absolute terms, but it is higher than expected at these stellar
luminosities \citep{tafel2010,cohenhuang2009}.  Meanwhile, as seen in
Fig. \ref{CL}, a  larger population may have been identified in lower mass
systems such as Ursa Major~II, Segue I, and Bo\"{o}tes. However they have lower metallicities,
with [Fe/H] between $-3$ and $-2.5$ and they have in the vast majority larger gravities
than the RGBs in the classical dSphs studied so far with \logg between 1.4 and 2
\citep{frebel2010b, norris2010b, lai2011}.

When it comes to small numbers, the sample selection criteria may seriously enter 
into play.  It is no secret that the detection of CEMPs in classical dwarfs is
random, i.e.  not particularly focused on C-abundances and 
the search for extremely metal-poor stars in those systems is notoriously difficult; 
 only nine known [Fe/H] $\leq -3$ stars have been discovered in Sculptor, one of the best studied
classical dSph. In comparison, dedicated works on the properties of EMPS and
the fraction of carbon stars in the Milky Way halo have been going on for many years with
targeted selections, allowing for example the creation of databases such as SAGA
\citep{suda2008} with $100$ [Fe/H] $\leq -3$ stars.

Restricting the comparison to the same stellar evolutionary range in Sculptor and
in the Milky Way halo, namely [Fe/H] $\leq -3$, \logg\ $\leq 1.6$, and
\Teff\ $\leq 4800K$, out of the compilation of \citet{placco2014}, there are three
major contributing samples: \citet{hollek2011} with 10 stars out of which 9 CEMP
stars defined as falling above the phenomenological line of \cite{aoki2007},
separating normal and carbon-rich stars (7 carbon rich stars with the Placco et al.
definition); \cite{yong2013} with 10 stars in total and 3 CEMP stars, and
\cite{barklem2005} with 13 stars out of which 2 CEMP stars (none with the Placco
et al. definition). This illustrates how sensitive the statistics are  to the selection of the
  sample, varying from 90\% to 15\% of stars with high
  carbonicity ([C/Fe] ratios, \citet{carollo2012}) in individual samples to 46\%
  when combining them all.

Another feature in calculating the statistics of carbonicity in different galaxies
is related to the spatial distribution. \citet{carollo2012} find that the fraction
of CEMP stars is larger by a factor of $\sim$1.5 between the inner and the outer halo for
$-3 \le$ [Fe/H] $\le -2.5 $, rising from $\sim$ 20\% to 30\%.
\citet{lee2013a} further notice that by restricting stars to within a $<$5kpc
distance from the Galactic mid-plane, the fraction of CEMP stars does not increase with
decreasing metallicity.  What part of the Milky Way halo should the samples in
dSphs, which are spatially randomly distributed, be compared with?

In conclusion, the comparison of the CEMP star fraction in dSphs and the Milky Way
is promising to unveil both the origin of carbon and the conditions of formation of
the different systems. However, it seems to require further investigation at least
along the RGBs in classical dwarfs, reaching larger gravities.

\subsection{The $\alpha$ elements}

\begin{figure}
\centering
\includegraphics[width=\hsize]{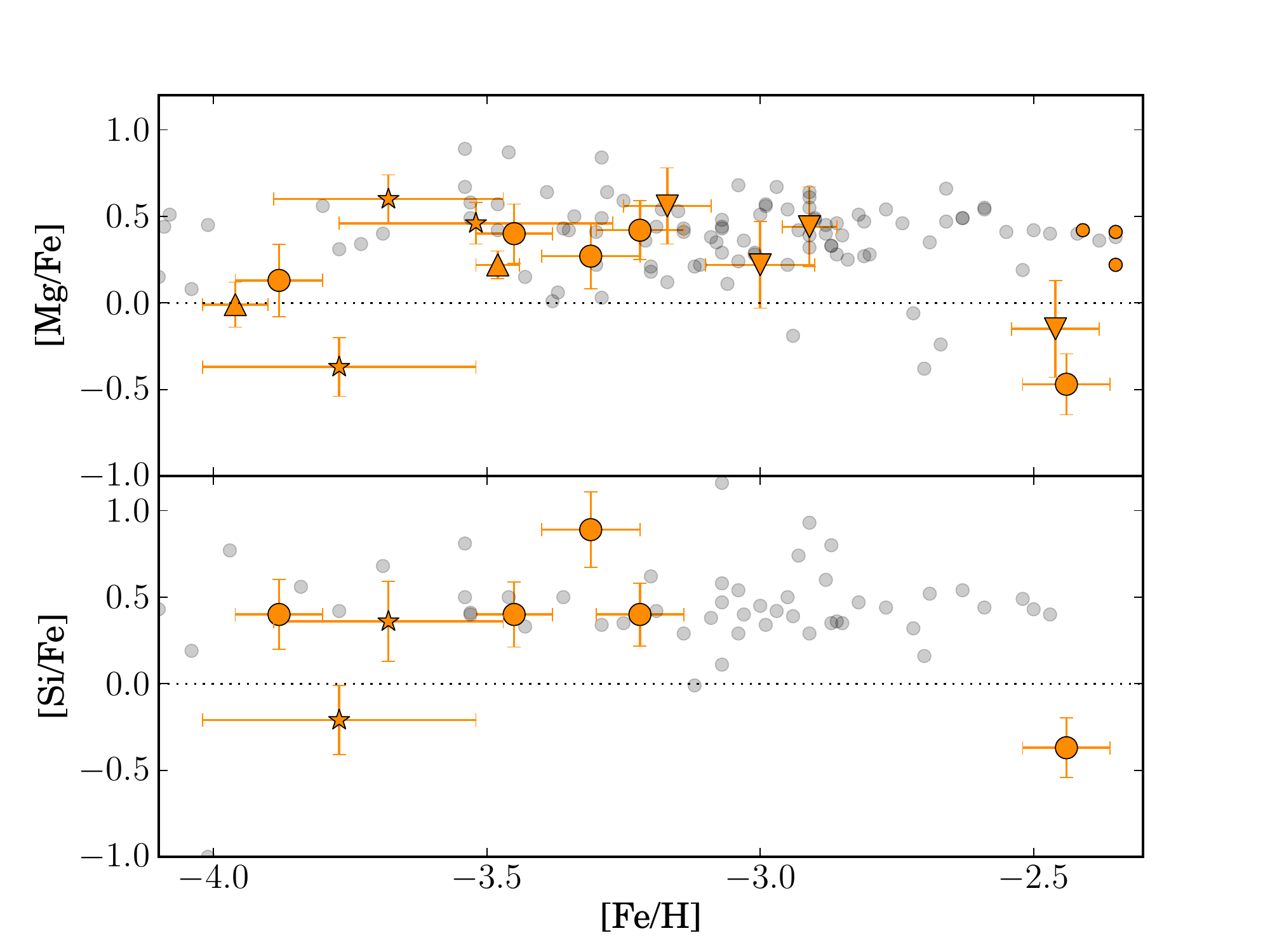}
\caption{Magnesium- and silicon-to-iron ratios as a function of [Fe/H] in Sculptor seen in orange,  compared to the Milky Way halo stars in grey\citep{honda2004,cayrel2004, spite2005,aoki2005, cohen2013, cohen2006,
  cohen2004, spite2006,aoki2007, lai2008, yong2013, ishigaki2013}.
We distinguish the new Sculptor sample presented in this paper by large circles with error bars. These error bars add in quadrature
the random and systematic uncertainties listed in Table~\ref{table:lines_abund} and Table~\ref{table:Tp_del_abun}. The sample of \cite{tafel2010} is shown with
upright triangles, while the \cite{starken2013} stars, which were not re$-$observed at high resolution, are displayed with inverted triangles. The
\cite{frebel2010a} and \cite{simon2015}  stars are indicated by a star.
}
\label{MgSi}
\end{figure}

\begin{figure}
\centering
\includegraphics[width=\hsize]{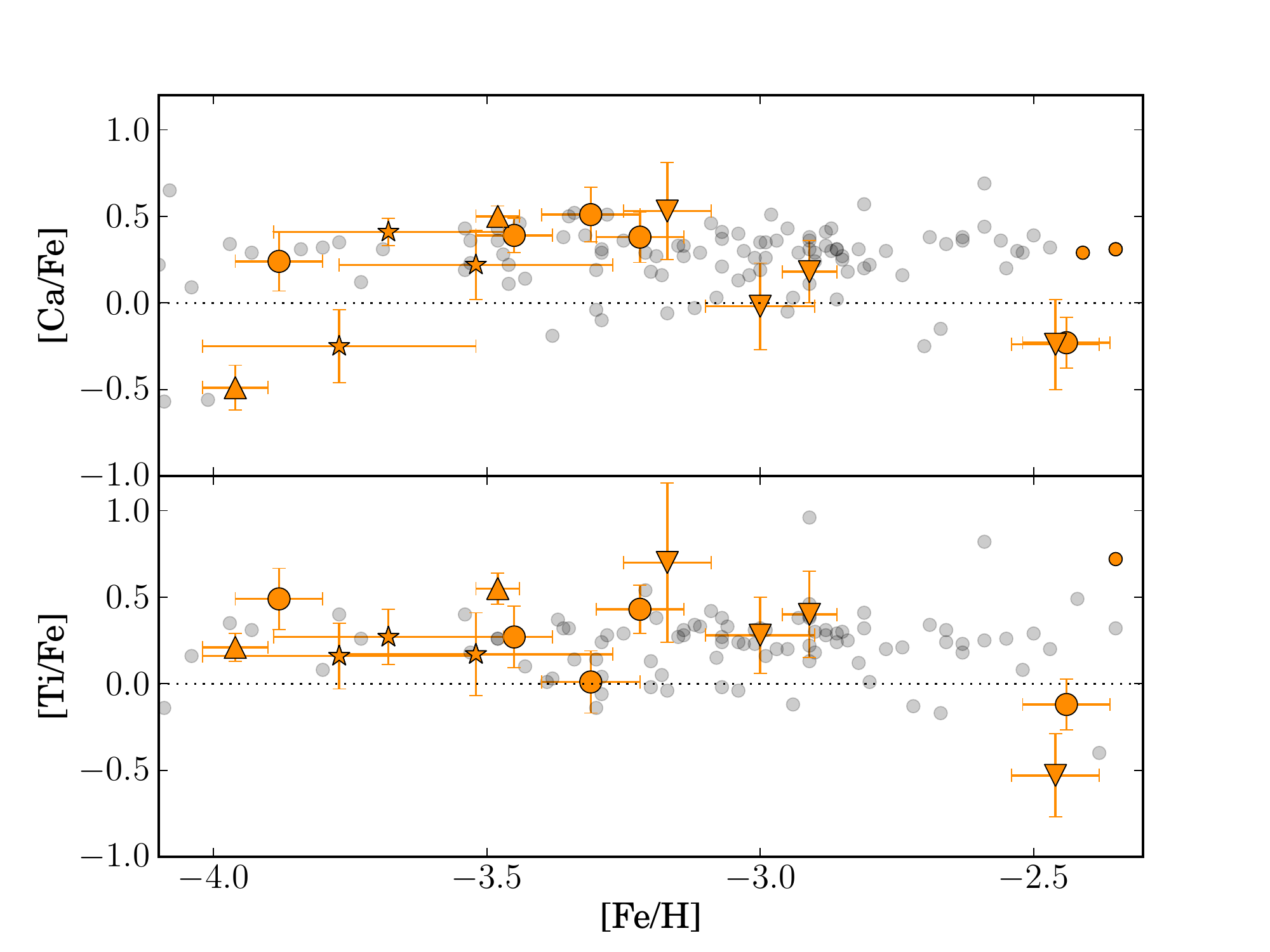}
\caption{Calcium- and titanium-to-iron ratios as a function of [Fe/H] in Sculptor seen in orange, compared to the Milky Way halo stars in grey.
The new Sculptor sample presented in this paper are shown  by large circles.}
\label{CaTi}
\end{figure}

Figures \ref{MgSi} and \ref{CaTi} display [Mg/Fe], [Si/Fe], [Ca/Fe], and [Ti/Fe] as a
function of [Fe/H]. For these elements as for the others, we present the results of
our analysis together with earlier works on Sculptor \cite{tafel2010, frebel2010a,
  starken2013}.

\subsubsection{Global features}

The bulk (80\%) of the metal-poor and extremely metal-poor stars observed so far in
Sculptor follow the main trend of the Milky Way halo stars for the four
$\alpha$-elements presented here, i.e. they mostly have supersolar abundance
ratios. The scatter of this plateau is also very similar in the two galaxies. There
are three outliers with subsolar ratios, ET0381, scl051\_05, and
scl\_11\_1\_4296. The star ET0381 at [Fe/H]$\sim -2.5$ is the only one with consistent low
ratios for the four elements and which can robustly be attributed to an interstellar
medium inhomogeneity, although  it is likely also the case of scl051\_05.  Indeed
\cite{starken2013} could not measure Si, but Ti, Mg, and Ca were securely derived from
2 to 5 lines.  The star scl\_11\_1\_4296 from \cite{simon2015} exhibits low
[$\alpha$/Fe] in three of the four elements, but a normal, Milky Way-like
[\ion{Ti}{II}/Fe]. Interestingly, the uncertainty on [Fe/H] for this star is similar to the underabundance in Mg, Si, and Ca. Moreover, its abundance of
\ion{Ti}{II} is derived from 11 lines as opposed to a single line in Si and Ca, and two
lines for Mg. This all calls for confirmation of these first determinations. Finally,
scl07-50 \citep{tafel2010} has a subsolar [Ca/Fe] ratio, but here again the Ca
abundance is derived from a single resonance line. A forthcoming paper dedicated to
accurate NLTE corrections will provide a better view, with a higher value, on the
intrinsic Ca abundance of this star \citep{mashonkina2015}.

In conclusion, so far robust evidence for the existence of pockets of chemical
inhomogeneity in the early days of Sculptor comes from two stars out of 14,
scl051\_05 and ET0381; the rest of the metal-poor population appears as homogeneous
as the Milky Way halo within the observational uncertainties. Indeed, a supersolar
plateau in [$\alpha$/Fe] is expected when stars form out of an interstellar medium in
which the ejecta of the massive stars, in numbers following a classical initial mass
function (IMF), are sufficiently well mixed \citep{pagel2009}.  This means that
whatever the process that led to this homogeneity, it was the same in Sculptor and
in the Milky Way halo, or at least, the processes at play were scalable to the
different sizes/masses of the systems.

\subsubsection{The origin of the outsiders}

There are two ways to produce the low [$\alpha$/Fe] ratios  observed for ET0381 and
scl051\_05. One is to increase [Fe/H] at a given $\alpha$-element abundance by the
ejecta of SNeIa and the second is to lower the abundance of $\alpha$-elements at a
fixed iron abundance, by varying the ratio between faint and massive SNeII. In the
following we investigate the two possibilities, concentrating on ET0381 for which we
have the most detailed chemical information:

$\bullet$ If  ET0381 were polluted by the ejecta of SNeIa, the second
nucleosynthetic signature of such an event would be a large production of
nickel. We can assume that before being polluted the ISM from which ET0381 arose
was of classical composition, i.e. on the [$\alpha$/Fe] plateau formed by the
SNeII products. Its initial metallicity [Fe/H]$_i$ would then be $-3.31 \pm 0.2$
(for [Mg/H]$=-2.91\pm 0.2$, i.e. [Mg/Fe]$\sim 0.4$), corresponding to
[Ni/H]$_i$=$-3.31 \pm 0.2$.  The W7 model of SNeIa nucleosynthesis, which is
widely used in galactic chemical evolution models \citep{Nomoto1984,iwamoto1999} and also in the work of \cite{Travaglio2005} varying the metallicity, predicts
$^{58}$Ni (a stable isotope of Ni) to be overproduced by a factor of 1 to $\sim$3
relative to iron, compared to the solar abundances.  If Ni were produced just
as much as Fe, the corresponding final abundance [Ni/H]=$-2.44$ would already
exceed the observations.  Chromium is also largely produced by SNeIa. Following
the same line of reasoning as for Ni, from an initial [Cr/H]$_i$=$-3.61 \pm 0.2$,
corresponding to the classical Milky Way halo level of a [Fe/H]=$-3.31 \pm 0.2$ star,
with a SNeI producing at least twice as much Cr than Fe compared to the solar
abundances, the final [Cr/H] should be $-2.44 \pm 0.2$ instead of $-2.75 \pm 0.1$.
Additionally, the knee of the [$\alpha$/Fe] versus the [Fe/H] relation in Sculptor is
found at [Fe/H]$>-2$ \citep{THT09}, hence at higher metallicity than the two
outliers shown here. This means that we are not witnessing a major onset of the
SNeIa explosions.  Therefore, ET0381 would need to form from an unusual type of
SNeIa, since the rest of the chemical evolution of Sculptor looks very chemically
homogeneous. The last piece of evidence comes from the existing observations
  of stars with metallicity larger than $-2$, i.e. past the metallicity at which
  the SNeIa are the major contributors to the chemical evolution of Sculptor
  \citep{THT09}. \citet{SVT03} measured the abundance of cobalt in four
  such stars, and they have [Co/Fe] close to or above solar, far above the present
  value for ET0381.

In summary, although there are no models of metal-poor SNeI that can definitely infer
the amount of iron peak elements they would release, for now there is a body of
corroborating evidence that a pollution by their sole ejecta is unlikely to explain
the chemical features of ET0381.

$\bullet$ The more massive the SNeII, the larger their yield of $\alpha$-elements.
Therefore, ET0381 (and possibly scl051\_05) could probably arise from a pocket of
interstellar medium that had not been polluted (or at least that was
under-polluted) by the most massive of the SNeII.

A quick calculation with the yields of \cite{tsujimoto1995} shows that missing the
ejecta of the stars more massive than 20M$_{\odot}$ leads to subsolar
[$\alpha$/Fe]. With such a truncated IMF, it takes only 30 Myr for a single stellar
population born from a gas clump 50 times more massive to reach [Fe/H] =$-2.5$ and
[Mg/Fe] =$-0.5$.  Although it is beyond the scope of the present paper to run full
and detailed simulations to understand how pockets of interstellar medium can have
discordant abundance ratios in an otherwise homogeneous galaxy, we ran a few tests
with the chemo-dynamical tree-SPH code GEAR \citep{revazjablonka2012}.  We followed
at very high spatial resolution (30 pc), the evolution of a
8$\times$10$^8$M$_{\odot}$ total mass system, forming stellar particles of 125
M$_{\odot}$ out of 500 M$_{\odot}$ gas particles. For the purpose of our test, we
switched off the explosions of the SNeIa and only considered the SNeII. We fixed the
star formation parameter c$_{\star}$ to 0.01 and the feedback efficiency to
$\epsilon$=0.5, allowing us to reproduce as closely as possible the properties of the
Local Group dSphs (Revaz et al., in prep.). Very low [$\alpha$/Fe] stars are indeed
created. They form in a dense gas shell expanding around the centre of the galaxy,
which is created by the explosion energy release from a strong star formation burst
at the origin of the first stars.  In this simulation, a 14M$_{\odot}$ SNeII polluted
a gas particle with initial [Fe/H] =$-4.2$.  This region receiving 0.6\% of the
ejected metals was then shifted to [Fe/H] =$-2.64$ and [Mg/Fe]=$-0.44$.  The rarity
of this event (5\% of the stellar population) and the fact that it falls naturally in
the rest of the evolution of the galaxy that we know lead us to favour this scenario
for ET0381, i.e. a lack of ejecta of the most massive SNeII.

\subsection{One odd-Z element: Scandium }

The NLTE corrections for scandium derived from \ion{Sc}{ii} are small, a few hundredths of a
dex \citep{zhang2008, zhang2014}, hence negligible, while the
corrections for Na and Al are much larger, of the order of a few tenths of a dex
\citep{andrievsky2007, andrievsky2008}, and will be published in
\citet{mashonkina2015}.

Scandium can be produced by a number of different channels in massive stars, but it is
not synthesized in SNeIa \citep{woosley2002}. Figure \ref{ScNi} shows how below
[Fe/H] $\sim -3$, the Sculptor EMPS beautifully follows the MW halo solar trend.
The only exception is ET0381, with [Sc/Fe] =$-0.43$, corroborating the hypothesis that
this star is  lacking products from SNeII ejecta.

\subsection{Iron peak-elements}

\begin{figure}
\centering
\includegraphics[width=\hsize]{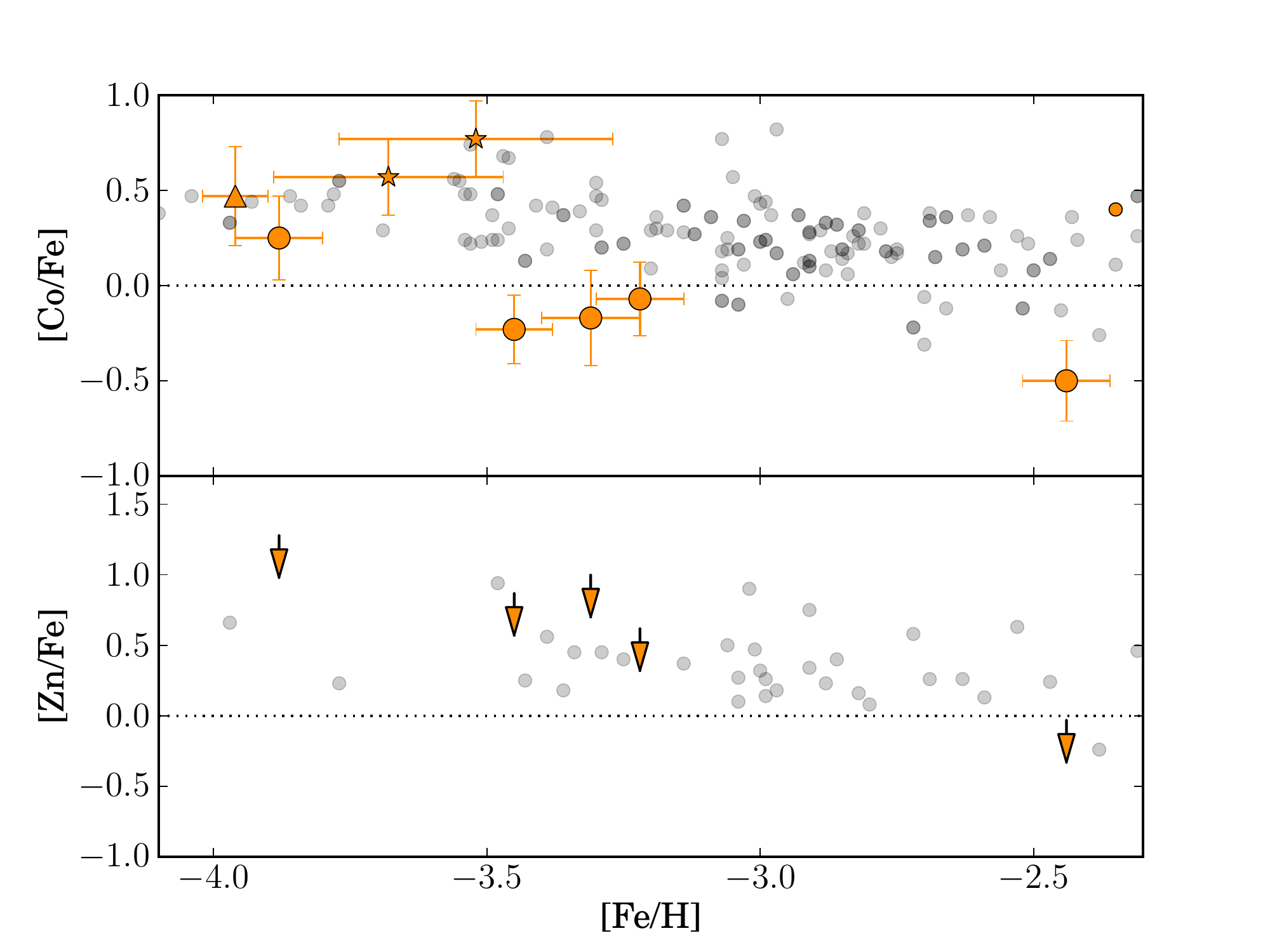}
\caption{Cobalt- and zinc-to-iron ratios as a function of [Fe/H] in Sculptor seen in orange,  compared to the Milky Way halo stars in grey.
The new Sculptor sample presented in this paper are shown  by large circles.}
\label{CoZn}
\end{figure}

\begin{figure}
\centering
\includegraphics[width=\hsize]{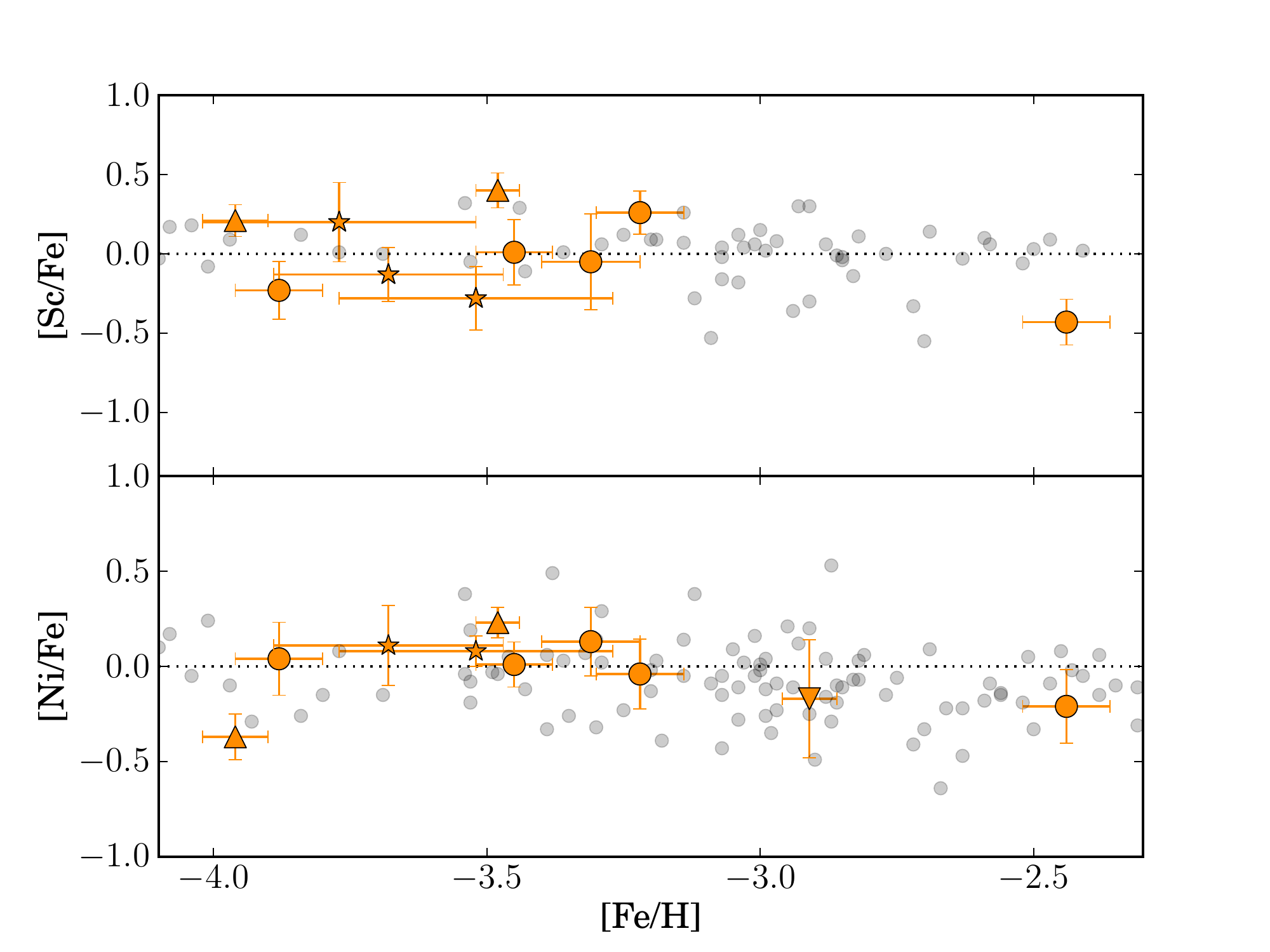}
\caption{Scandium- and nickel-to-iron ratios as a function of [Fe/H] in Sculptor seen in orange,  compared to the Milky Way halo stars  in grey.
The new Sculptor sample presented in this paper are shown  by large circles.}
\label{ScNi}
\end{figure}

\begin{figure}
\centering
\includegraphics[width=\hsize]{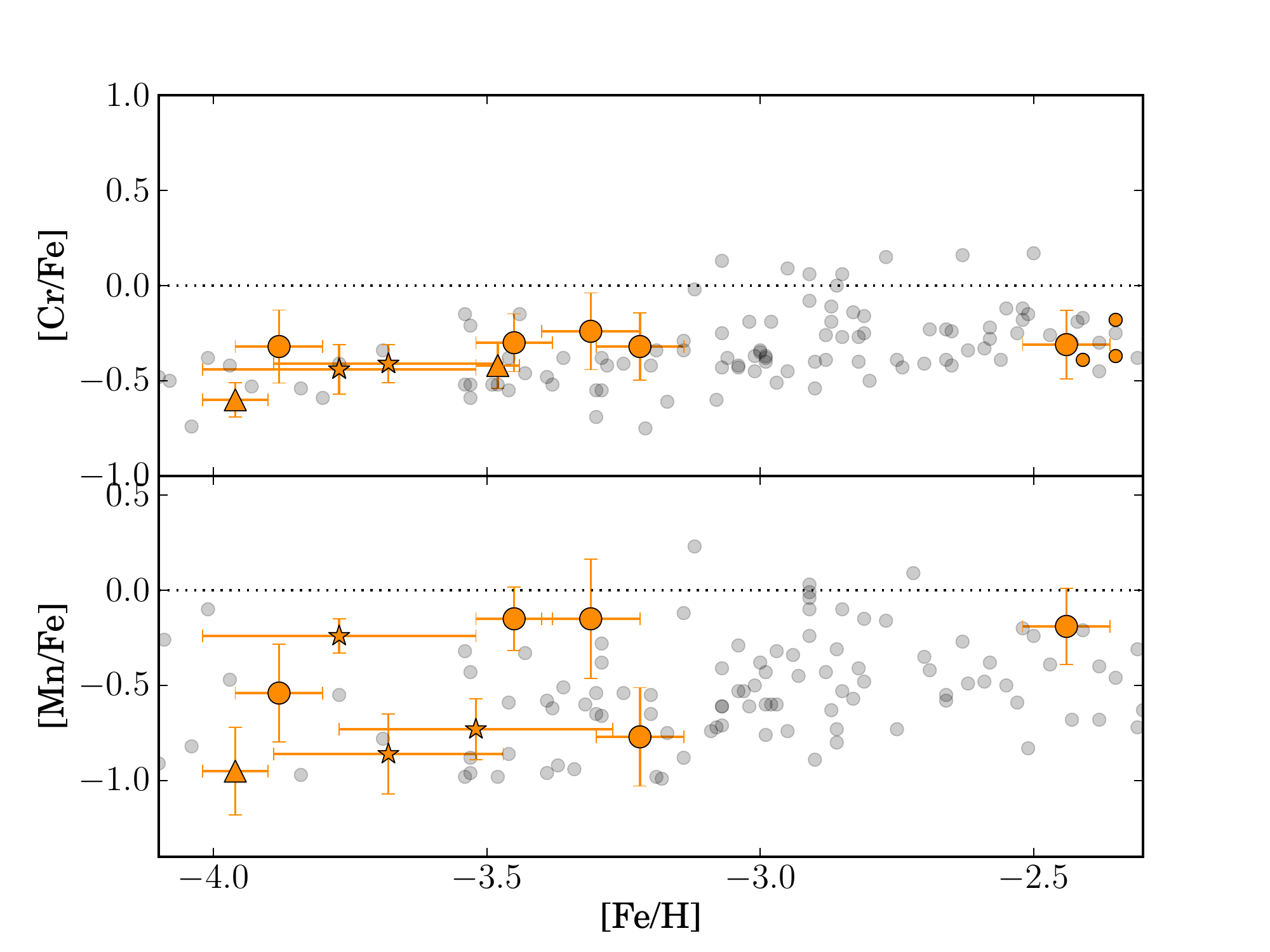}
\caption{Chromium- and manganese-to-iron ratios as a function of [Fe/H] in Sculptor seen in orange,  compared to the Milky Way halo stars  in grey.
The new Sculptor sample presented in this paper are shown  by large circles.}
\label{CrMn}
\end{figure}

Figures \ref{CoZn}, \ref{ScNi}, and \ref{CrMn} present the elements produced by explosive
nucleosynthesis, and specifically the behaviour  of the elemental abundance ratios relative
to iron of nickel, cobalt, chromium, and manganese as a function of
metallicity.

In the Milky Way, the [Ni/Fe] abundance ratio is roughly constant in stars of very
different metallicities. This is generally understood as the fact that Ni is produced
abundantly by both complete and incomplete Si burning \citep{umeda2002}, and the
present dataset is no exception to this rule.

The other elements are likely more informative, because they are produced in two
distinct regions characterized by the peak temperature of the shock material.
Above $5x10^9$K material undergoes complete Si burning and its products include Co
and Zn, while at lower temperature incomplete Si burning takes place and after
decay produces Cr and Mn \citep[e.g.][and references therein]{umeda2002}. The
mass cut, which divides the ejecta and the compact remnant, is located close to the
border of complete and incomplete Si-burning regions.  The large [Zn/Fe] and
[Co/Fe] values at low metallicity in the Milky Way halo stars are a challenge to
nucleosynthesis models and the exact relations between the explosion energy,
the mass of the stars and the fallback/mixing processes are still much discussed
by the experts \citep[][]{nakamura1999, umeda2005, limongi2012}. Milky Way
halo stars do not seem discriminant enough to distinguish between the different
scenarii.  Our new sample of EMPs may  help shed light on these questions,
in particular by constraining the role of the progenitor mass.

With its low [Co/Fe] and [Zn/Fe], ET0381 stands out clearly  from the Galactic trends,
while it follows perfectly the Milky Way halo stars in [Cr/Fe] and [Mn/Fe].  
This is again consistent with the assumption that this star is missing the product
of the high-mass tail of the SNeII, i.e. the major producers of Co and Zn, and
confirms that the depth of the mass cut or the explosion energy does vary with
the SNeII progenitor mass, the most massive ones having deeper mass cuts or more
energetic explosions. It is not possible at this stage to distinguish between the depth
of the mass cut and energy: the complete Si-burning region is enlarged in both cases;
increasing the energy produces effects similar to making the mass cut deeper without
changing the mass coordinate of the mass cut \citep{umeda2005}. Meanwhile the outer
Si-incomplete burning regions are most insensitive to the cut or energy, explaining
why there is no signature of the specificity of ET0381 in Mn and Cr.  The fact that
one can estimate  which stellar mass range is missing in
the composition of ET0381 from the $\alpha$-elements makes this star a  unique calibrator for the models of
nucleosynthesis.

Two other stars, scl002\_06 and scl074\_02, lie slightly below the Galactic halo
star trends in Co. However, their case is different from that of ET0381; they are not simultaneously depleted in Zn. As will be discussed in
\citet{mashonkina2015} the magnitude of the NLTE correction for Co in these two
stars is $\sim$0.2 dex, reducing significantly the gap between them and the Milky
Way halo, without fully withdrawing their difference (1$\sigma$). These stars would have a solar [Co/Fe] at [Fe/H] =$-3.5$, while this level is
reached at [Fe/H] =$-3$ in the Milky Way halo.  We could be witnessing a
consequence of the difference in star formation history between the two
galaxies. In a similar way to the knee in [$\alpha$/Fe], which  appears at lower
metallicity in galaxies forming stars less efficiently, the decline in [Co/Fe] is
seen at lower metallicity in Sculptor than in the Milky Way.  This hypothesis
needs further support from additional observations in the [Fe/H] =$-3.5$ to $-2.5$
range.

\subsection{Neutron capture elements}
\label{neutron}

\subsubsection{General trends}

\begin{figure}
\centering
\includegraphics[width=\hsize]{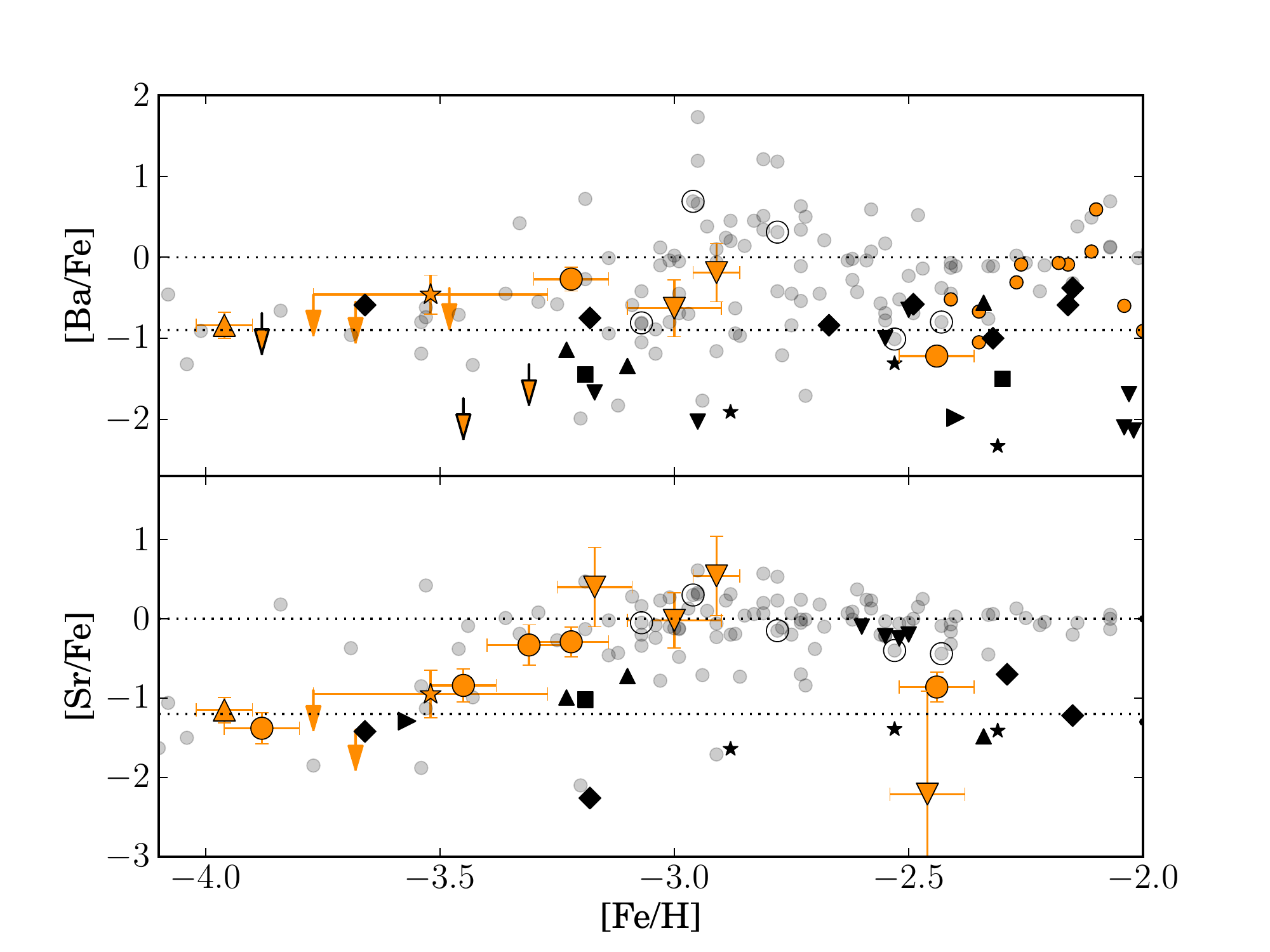}
\caption{Barium- and strontium-to-iron ratios as a function of [Fe/H] in Sculptor seen in orange,  compared to the Milky Way halo stars in grey. The open
circles correspond to HE1219-0312, HD140283, HD122563, HD88609, and HE2327-5642, five Milky Way halo stars for which the odd-$A$ fraction could be estimated. 
The ultra$-$faint dwarf spheroidal galaxies are displayed with black symbols.  Ursa Major~II is identified with triangles, Coma Berenices with stars, Leo IV  with squares, Hercules  with pointing down triangles, Segue I  with pointing right triangles, and Bo\"{o}tes with diamonds. The full references 
  are given in Sect.~\ref{symbols}.
}
\label{BaSr}
\end{figure}

Figure \ref{BaSr} characterizes the two neutron capture elements that we could
measure in our stars, barium and strontium, as a function of the iron abundance,
while Fig.\ref{SrBaBaH} indicates the positions of the two dominating productions of
Ba and Sr at different stages of the galaxy chemical evolution, the weak and main
r-process, following the observations and definitions of \cite{francois2007}. All
Sculptor stars at [Fe/H]$>-3.5$ with Sr and Ba measurements fall in the weak
r-process regime and so would the rest of stars with upper limits only in Ba.  Below
[Fe/H]$\sim-3.5$, the stars in Sculptor follow the Milky Way halo trend in [Ba/Fe]
and [Sr/Fe], with [Sr/Ba] ratios which can be low even at [Ba/H]$\sim -4$, again 
similar to the value measured in our Galaxy.

Figure \ref{BaSr} suggests that the first channel for the production of Ba and Sr is
in place early ($-4 \leq$ [Fe/H] $\leq -3.5$) and in all galaxies whatever their
stellar masses. Core collapse supernovae would therefore make a natural channel of
production, as we know from the other class of chemical elements that they do explode
\citep{woosley1994, qian2007}. However, this channel would only be applicable to the
abundance floor in heavy (Ba) and light (Sr) elements.  It is interesting to note
that ET0381, which is most likely missing some of highest mass SNeII ejecta, is not
particularly low in [Sr/H] and [Ba/H]. Its level of enrichment is comparable to the
other Sculptor stars in the weak-process regime. This is a clear sign that
medium-mass SNeII contribute to the production of neutron capture elements and
particularly to nucleosynthesis of the light elements. Along this line Draco119, a
[Fe/H]$=-2.9$ star in Draco which is shown by \citet{fulbright2004} to be very
depleted both in Ba and Sr, is also suspected of having missed the ejecta of the
low-mass tail of the SNeII.

As mentioned above, excluding ET0381 and scl051\_05 given their global peculiar
chemical patterns, both the effective measurements and the upper limits in Sculptor
indicate that [Ba/Fe] and [Sr/Fe] increase from [Fe/H] =$-3.5$ with increasing
[Fe/H], just as the Galactic halo does.  This is in sharp contrast with the behaviour
of the neutron-capture elements in the UFDs.  With the exception of three stars in
Hercules \citep{francois2012}, the abundance of the neutron-capture elements does not
rise in the UFDs. Instead it scatters around a mean floor level of $\sim$$-1.3$ in
[Sr/Fe] and $-0.9$ in [Ba/Fe]. The mass of Hercules is highly uncertain; given that
it is considerably elongated \citep{coleman2007} hence potentially stripped, it was
likely more massive in the past than is measured today.

\subsubsection{The role of galaxy mass}

Mass is the major structural difference between Sculptor (or classical dSphs) and the
UFDs: while Sculptor has a $V$-magnitude M$_\mathrm{V}$$\sim-11.2$, and a velocity
dispersion $\sigma \sim$9 - 10 km/s \citep{BHT08,walker2009}, the UFDs have much
smaller masses with $\sigma$ between $\sim$3 and 6 km/s and have a sparse stellar
population, M$_\mathrm{V}<-4$ mag \citep{simongeha2007,geha2009}. The probability of
producing Sr and Ba seems to correlate with the number of massive stars formed in the
systems. \cite{tafel2010} also reported normal [Sr/Fe] and [Ba/Fe] ratios in Sextans,
supporting this idea.  In Fig. \ref{SrBaBaH} galaxies more massive than Carina (i.e. known to have an efficient early period of star
formation), follow the trend defined by the MW halo stars. This suggests that the
main nucleosynthesis source of both the light and the heavy r-process elements is
linked to the galaxy stellar mass content, in the sense that there is elemental
enrichment only above a given mass threshold.  Independently, \citet{tsujimoto2014a,
  tsujimoto2014b} recently promoted the neutron star mergers as low-probability
events satisfying the above constrains. This is a very promising hypothesis also
because it seems that the SNeII do not  synthesize the heavy r-process elements
\citep{wanajo2013}, while the neutron star mergers are successful in producing
both the heavy and light r-process elements \citep{wanajo2014,just2015}.

\subsubsection{Coupling different channels of production?}

\begin{figure}
\centering
\includegraphics[width=\hsize]{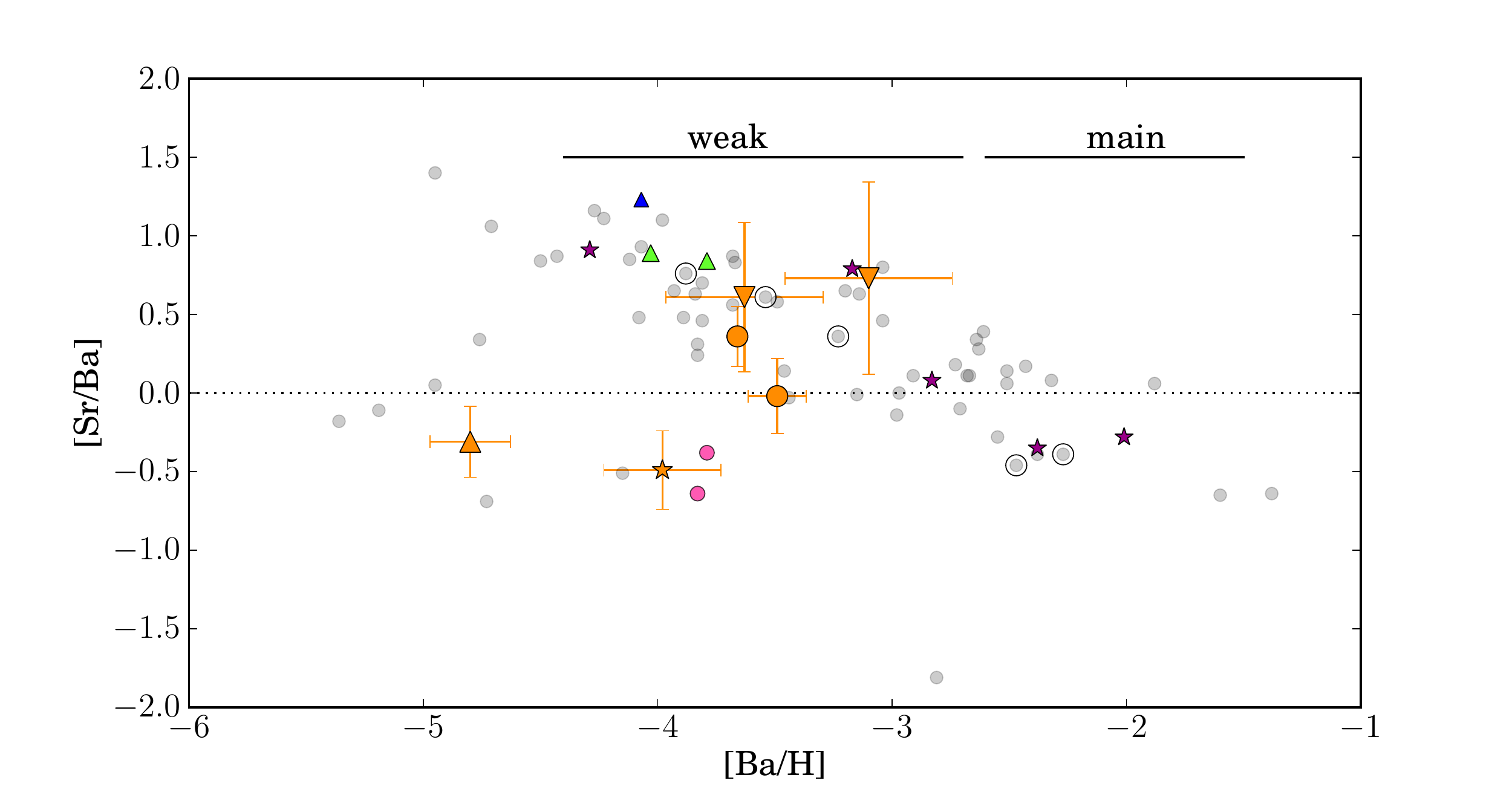}
\caption{Barium-to-strontium ratio as a function of [Ba/H] in Sculptor seen in orange,  compared to the Milky Way halo stars presented  in grey. The open
circles correspond to HE1219-0312, HD140283, HD122563, HD88609, and HE2327-5642, five Milky Way halo stars for which the odd-$A$ fraction could be estimated. 
Sextans is shown with green triangles, Fornax with a blue triangle. The Ursa Minor population is shown with
purple stars while the two Carina  stars  are seen in pink circles. The full references   are given in Sect.~\ref{symbols}. }
\label{SrBaBaH}
\end{figure}

Our dataset provides an additional piece of information, which should help the
  identification of the nucleoynthesis site of Ba and Sr.  We indicate in Figures
  \ref{BaSr} and \ref{SrBaBaH} the position of five [Fe/H] $<-2.5$ Milky Way halo
  stars for which the Ba odd-$A$ isotope fraction has been derived so far.  They all
  have [Eu/Ba] between 0.48 and 0.79, classically interpreted as the signature of an
  r-process.  Three stars have measured low or moderate $f_\mathrm{odd}$: HD88609
  \citep[$f_\mathrm{odd}$ =$-0.02\pm$0.09,][]{gallagher2012}, HD122563
  \citep[$f_\mathrm{odd}$ =0.22 $\pm$0.15,][]{mashonkina2008}, and HD140283
  \citep[][with $f_\mathrm{odd}$ =0.15 $\pm$0.12 and $f_\mathrm{odd}$ =0.02
    $\pm$0.06, respectively]{collet2009,gallagher2012}.  The exact value of
  $f_\mathrm{odd}$ of HD140283 continues to foster discussion in the
  literature. Noticeably, the various determinations are all based on the the shape of
  the \ion{Ba}{ii} 4554\,\AA\ line, but the results oscillate between a minimum value
  of $f_\mathrm{odd}$ at 0.02 and a  maximum at 0.38 \citep{Magain1995,
    2002MNRAS.335..325L, gallagher2015}, still below the pure r-process threshold.
  Two other stars, HE2327-5642 \citep[$f_\mathrm{odd}$=0.5,][]{mashonkina2010} and now
  HE1219-0312 from this study ($f_\mathrm{odd}$ $\ge 0.46$), have high
  $f_\mathrm{odd}$ signature of a pure r-process. They also have the largest Ba
abundances. The other three stars have low [Ba/Fe], corresponding to the [Ba/Fe]
floor level as defined by the [Fe/H]$\in$[$-4$,$-3$] stars.  In Figure \ref{SrBaBaH},
HD122563, HD140283, and HD88609 fall in the so-called weak r-process regime and so do
ET0381 and scl\_03\_059, while HE1219-0312 and HE2327-5642 are clearly in the main
regime.  Hence, the value of $f_\mathrm{odd}$ could be linked to the nucleosynthetic
origin of the neutron capture elements, which may  be different in the two
regimes.

Adopting the nucleosynthesis prescriptions of \citet{frischknecht2012},
\citet{cescutti2013} proposed that stars with high [Sr/Ba] abundances are polluted
by the s-process in massive (10-40 M\sun) rapidly rotating stars, and predicted
that they should have low values of the odd/even isotopes of Ba. This matches our results
well.  There is one caveat, however (also seen in
\cite{cescutti2014}): the stars with low $f_\mathrm{odd}$ also have very low model
      [Ba/Fe] and [Sr/Fe], while in the Sculptor and Milky Way halos stars, if
      [Ba/Fe] can be low, [Sr/Fe] is already close to solar. Leaving room for
      future improvement and refinement of the nucleosynthesis spinstar models, one
      can still seriously consider the hypothesis that part of the neutron capture
      elements are produced through the s-process channel in massive stars.  This
      would even ease the understanding of the low [Ba/Fe] and [Sr/Fe] in
      UFDs. Indeed, these stars exhibit a normal [$\alpha$/Fe] plateau, hence they do not
      miss the high-mass part of the IMF; moreover, most of them have high
      [Sr/Ba]. This early channel of production of the neutron capture elements
      could be coupled with rarer events such as neutron star mergers as suggested
      above, which would only be able to enrich massive galaxies.  Models of
      pure r-process should probably also be investigated in order to see whether
      they can produce low $f_\mathrm{odd}$. To our knowledge this has not yet
      been done, or  published.

\subsection{To be or not to be pair-instable}

\begin{figure}
\centering
\includegraphics[width=\hsize]{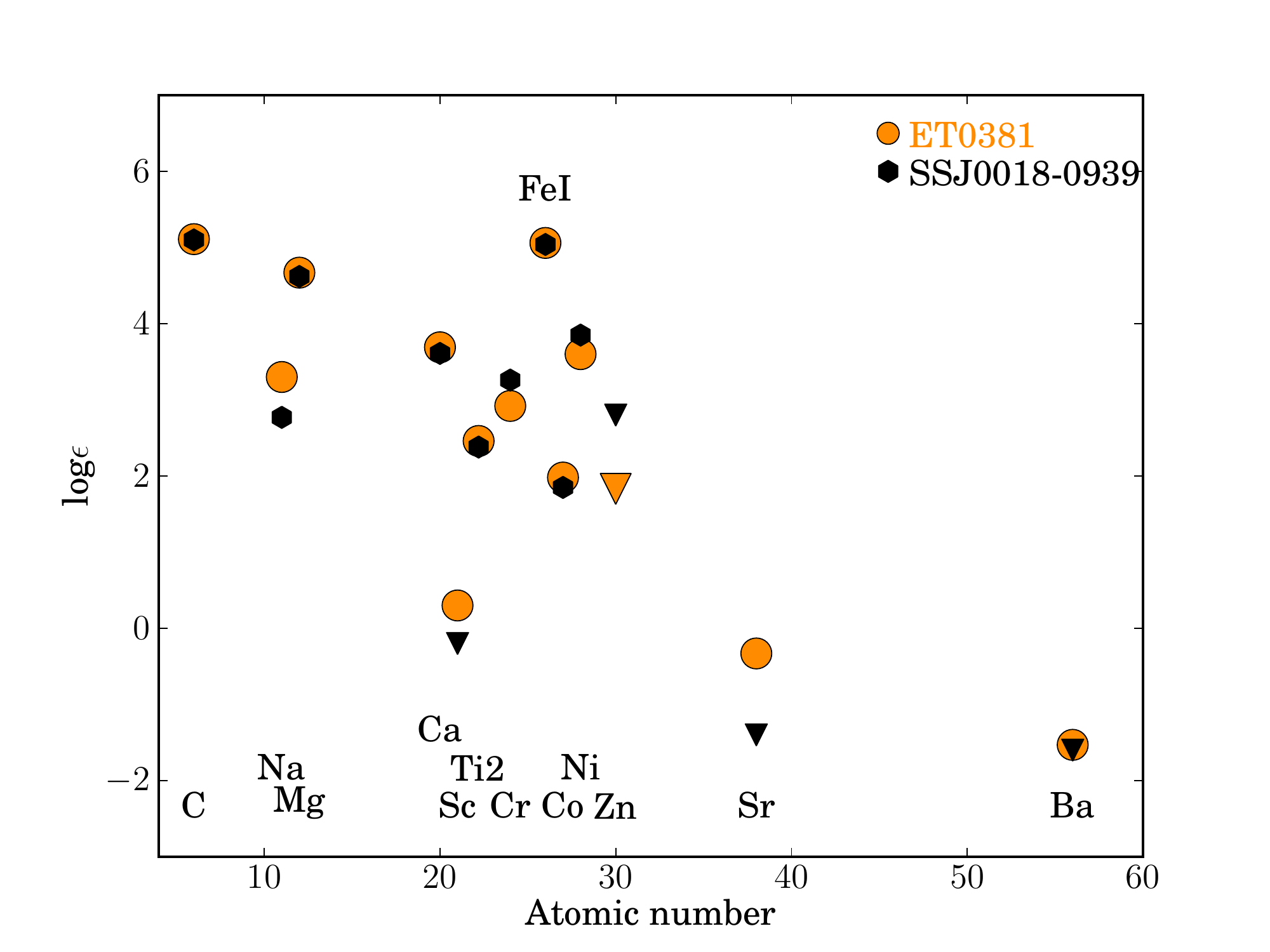}
\caption{Comparison of the elemental abundances of ET0381 in the Sculptor dSph and  SDSS J001820.5-093939.2 in the Milky Way halo. The upper limits are identified
by triangles.}
\label{pair}
\end{figure}

The low [$\alpha$/Fe] ratio and large contrasts between the abundances of odd and
even element pairs of the Milky Way halo star, SDSS J001820.5-093939.2, a cool
main-sequence star, led \cite{aoki2014} to consider the hypothesis that this star was
holding the pattern records of a pair-instability supernova (PISN). Figure \ref{pair}
compares the abundances of SDSS J001820.5-093939.2 and ET0381 in log$\epsilon$
units. The similarity in all measured elements between the two stars is striking and
raises the possibility of a common origin. There are a number of low [$\alpha$/Fe] stars
known \citep{ivans2003,venn2012}, but none shares so many identical features.

\cite{aoki2014} rightly pointed out that very low carbon abundances are found in
highly evolved red giants and are often interpreted as the result of internal
extra-mixing. According to \citet{placco2014}, the original carbon abundance
  of ET0381 should be about 0.7dex higher than measured. The rest of the chemical
  elements on which we can base the comparison with SDSS J001820.5-093939.2 are
  not prone to modifications due to mixing.

There is a major obstacle to ET0381 tracing the ejecta of a PISN. A simple
calculation shows that the explosion of a $\sim$200$\times 10^{51}$erg event would
exceed the binding energy of the gas in a virialized $\sim$5$\times 10^8$
M$_{\odot}$ dwarf system, such as Sculptor, assuming a Plummer sphere profile, in
agreement with the more sophisticated simulations of \cite{wada2003}. Of course,
the exact feedback energy that a system can sustain depends on the galaxy mass and
profile: the more concentrated the profile, the stronger the resistance of the
galaxy. However, even before considering the extreme feedback limit when the gas
is ejected to infinity, one faces the fact that the huge amount of feedback energy
heats the gas to a temperature above 10$^6$K, and makes it expand, strongly
decreasing its density. The gas cooling time becomes very long, moving from the Myr to
Gyr scale. Therefore, the time necessary for the galaxy to be able to form stars
again is clearly not negligible, whereas there is no sign in the Sculptor star
formation history of any quiescent period.  Along a similar line of argument,
\cite{revaz2009} demonstrated that the chemical homogeneity of the Sculptor and
Fornax dSphs  strongly constrains the supernova effective feedback energy,
setting it up to a low value.

In conclusion, if the peculiar chemical abundance pattern of SDSS J001820.5-093939.2
has the same origin as that of ET0381, it could be due to a depletion in massive
SNeII ejecta rather than to the signature of a PISN. This could also alleviate the
much larger Na, Al, V, and Mn in SDSS J001820.5-093939.2 than in the PISN synthesis
models \cite{aoki2014}.

\section{Summary}
\label{conclusion}

We presented the high-resolution spectroscopic analysis of five very metal-poor stars
in the Milky Way dSph satellite, Sculptor. This doubles the number of stars in this
metallicity range with comparable observations.  The abundances of 16 elements could
be derived: \ion{Fe}{i}, \ion{Fe}{ii}, C, \ion{Na}{i}, \ion{Mg}{i}, \ion{Al}{i},
\ion{Si}{i}, \ion{Ca}{i}, \ion{Sc}{ii}, \ion{Ti}{i}, \ion{Ti}{ii},
\ion{Cr}{i}, \ion{Mn}{i}, \ion{Co}{i}, \ion{Ni}{i}, \ion{Sr}{i}, and \ion{Ba}{ii}. Upper
limits could be estimated for three more elements: \ion{Zn}{i}, \ion{Y}{ii}, and
\ion{Eu}{ii}.

In combination with previous works the low-metallicity tail of the early generation
of stars in Sculptor could be better characterized, with consequences for our
understanding of galaxy evolution and stellar nucleosynthesis. Our main results can
be summarized as follows:

\begin{itemize}

\item The bulk (80\%) of the metal-poor and extremely metal-poor
  stars observed so far in Sculptor unambiguously follow the main trend of the
  Milky Way halo stars for the $\alpha$-elements presented here.  Both populations are
  also similar in iron-peak and neutron capture elements. This is expected when stars
  form out of an interstellar medium in which the ejecta of the massive stars, in
  numbers following a classical initial mass function, are sufficiently well mixed.
  This implies that the early conditions of star formation were the same in Sculptor
  and in the Milky Way halo, or at least, that the processes at play were scalable to
  the different sizes/masses of the systems.

\item Despite overall chemical homogeneity, our dataset reveals one new star, ET0381,
  at [Fe/H] $=-2.5$ which is remarkably poor in $\alpha$ and iron-peak elements for
  its metallicity. From nucleosynthesis arguments and with the help of a few simple
  chemo-dynamical simulation, we conjecture that ET0381 arose from a pocket of
  interstellar medium missing the ejecta of the most massive Type II supernovae.

\item The analysis of the iron-peak element abundances of ET0381 supports the
  theoretical predictions that the depth of the mass cut and/or the explosion energy
  vary with the SNeII progenitor mass, and that Co and Zn are largely produced by the
  high-mass tail of the massive stars. The fact that one can estimate from the
  $\alpha$-elements which SNII mass range is missing in the composition of ET0381
  makes this star a unique calibrator for the models of nucleosynthesis.

\item Our analysis brings important clues on the nucleosynthetic site of the neutron
  capture elements: \\ {\it i)} We find that the gradual enrichment in barium and
  strontium of the Sculptor metal-poor stars closely follows the evolution of the MW
  halo, contrary to the stellar population in ultra-faint dwarfs. This provides
  further evidence that the mass of a galaxy is an important driver of its chemical
  evolution and more specifically of its ability to produce neutron capture
  elements. \\ {\it ii)} The Sculptor $-3.5 <$[Fe/H]$<-2.5$ stars for which both Sr
  and Ba has been measured so far, fall in the so-called weak r-process regime, like
  stars in the other massive classical dwarfs.  \\ {\it iii)} The comparison of the
  abundances derived from the \ion{Ba}{ii} subordinate and 4934\,{\AA} lines reveals
  that their agreement can only be achieved for a solar Ba isotope mixture, meaning a
  low odd-to-even isotope abundance ratios. Further comparison with a set of Milky
  Way halo stars, suggests that $f_\mathrm{odd}$ is linked to the origin of the
  neutron capture elements. Low $f_\mathrm{odd}$ are predicted by the models of
  spinstar producing Ba and Sr through the s-process, but some of their predictions
  are not corroborated by our observations. We call for refined models and
  investigation on whether pure r-process could generates low $f_\mathrm{odd}$ as
  well.  \\ {\it iv)} We find evidence that the medium-mass SNeII play a role in the
  production of the light neutron capture elements, as represented by Sr. \\ {\it v)}
  We postulate a double (at least) origin of the neutron capture elements.  One is
  most probably linked to the massive stars that generate the abundance floor; the
  other could be related to rare events correlating with the stellar mass of the
  galaxies, such as the neutron star mergers.

\item A comparison between ET0381 and the Milky Way halo star SDSS J001820.5-093939.2
  reveals striking similarities and suggests a common origin. It is highly improbable
  that ET0381 could hold the pattern records of a pair-instability supernova, as was
  proposed for SDSS J001820.5-093939.2, because the Sculptor dSph would be disrupted
  by the enormous energy release at the time of the SNeII explosions. Even without
  disruption, a long period of quiescence in its star formation, imposed by the long
  gas cooling time, should be observed, but it is not. We suggest that SDSS
  J001820.5-093939.2, like ET0381, is missing the ejecta of the most massive SNeII
  and that the signature of a PISN has yet to be found.

\end{itemize}

\begin{acknowledgements}
  Pascale Jablonka and Pierre North gratefully acknowledge financial support from
  the Swiss National Science Foundation.  Lyudmila Mashonkina is supported by the
  Russian Foundation for Basic Research (grant 14-02-91153).  Else Starkenburg
  gratefully acknowledges funding through the CIFAR Global Scholar network.  This
  work also benefited from the International Space Science Institute (ISSI) in
  Bern, thanks to the funding of the team ``The first stars in dwarf
  galaxies''. We thank the VISTA commissioning team for providing $J$, $K_s$
  photometry which was observed during the commissioning process. We thank
  Prof. Bertrand Plez for his help in the use of \verb+turbospectrum+ We enjoyed
  lively and fruitful discussions with M. Limongi, G. Meynet, and S. Wanajo.
\end{acknowledgements}

\bibliographystyle{aa} 
\bibliography{scl_emps}

\newpage

\begin{table*}[h]
\caption{Journal of the observations. The successive columns provide the target
  coordinates, wavelength ranges of the spectra, as well as their corresponding
  signal-to-noise ratios per pixel and radial velocities from this work. The last column
  gives the metallicities estimated from low-resolution spectroscopy of the
  near-infrared calcium triplet (CaT).}
\label{table:journal_obs}
\centering
\begin{tabular}{lrrrrrrrrrr} \hline \hline
\multicolumn{1}{c}{ID}&\multicolumn{1}{c}{$\alpha$ (J2000)}&\multicolumn{1}{c}{$\delta$ (J2000)}&
\multicolumn{1}{c}{$t_\mathrm{exp}$}&\multicolumn{1}{c}{$\lambda$ range}&
\multicolumn{1}{c}{S/N}&\multicolumn{1}{c}{V$_\mathrm{rad}$}&\multicolumn{1}{c}{$\langle v_\mathrm{rad}\rangle$}&\multicolumn{1}{c}{[Fe/H]$_\mathrm{CaT}$}\\
\multicolumn{1}{c}{[Fe/H]$_\mathrm{CaT}$} \\
&\multicolumn{1}{c}{[h:mn:s]}&\multicolumn{1}{c}{[\degr :':'']}&
\multicolumn{1}{c}{[s]}&\multicolumn{1}{c}{[\AA]}&&\multicolumn{1}{c}{[km\,s$^{-1}$]}&
\multicolumn{1}{c}{[km\,s$^{-1}$]}& \\ \hline
ET0381    &$01:00:12.94$&$-33:42:03.8$&$6000$&$3548-4525$&$10,15$&$102.64$&&\\
          &$           $&$           $&$5900$&$4786-5760$&$30,40$&$102.56$&        &     \\
          &$           $&$           $&$5000$&$5837-6807$&$60,67$&$102.75$&$102.65$&$-2.75$\\
\hline
scl002\_06 &$01:01:26.74$&$-33:02:59.8$&$3000$&$3593-4524$&$10,14$&$117.31$&&\\
          &$           $&$           $&$3000$&$4787-5760$&$40,50$&$117.16$&&\\
          &$           $&$           $&$3000$&$5836-6807$&$60,60$&$117.03$&$117.17$&$-3.04$\\ \hline
scl\_03\_059&$01:01:22.23$&$-33:46:21.3$&$5920$&$3626-4523$&$15,18$&$ 87.59$&&\\
          &$           $&$           $&$5000$&$4786-5759$&$38,50$&$ 87.83$&&\\
          &$           $&$           $&$6000$&$5835-6806$&$60,70$&$ 88.16$&$ 87.86$&$-2.82$\\ \hline
scl031\_11 &$00:57:10.21$&$-33:28:35.7$&$6000$&$3656-4524$&$12,17$&$119.71$&&\\
          &$           $&$           $&$6300$&$4786-5758$&$37,44$&$120.54$&&\\
          &$           $&$           $&$5900$&$5835-6806$&$65,67$&$119.07$&$119.77$&$-3.61$\\
\hline
scl074\_02 &$00:57:34.84$&$-33:39:45.6$&$3800$&$3674-4524$&$ 9,12$&$130.40$&&\\
          &$           $&$           $&$6000$&$4727-5804$&$25,30$&$130.51$&&\\
          &$           $&$           $&$6000$&$5818-6835$&$45,45$&$131.52$&$130.81$&$-3.02$\\
\hline
\end{tabular}
\end{table*}

\begin{table*}[h]
\caption{Line parameters, observed equivalent widths, and abundances of the five EMP stars of the Sculptor dSph galaxy.}
\label{table:lines_abund}
\centering
\scriptsize
\begin{tabular}{lrrr|rr|rr|rr|rr|rr} \hline \hline
El.&\multicolumn{1}{c}{$\lambda$}&$\chi_\mathrm{ex}$&$\log(gf)$&\multicolumn{10}{|c}{EQW$\pm\Delta$EQW~~[m\AA]~~~$\log\varepsilon(X)+12$~~[dex]} \\
&&&&\multicolumn{2}{|c}{ET0381}&\multicolumn{2}{|c}{scl002\_06}&\multicolumn{2}{|c}{scl\_03\_059}&\multicolumn{2}{|c}{scl031\_11}&\multicolumn{2}{|c}{scl074\_02}\\
 \hline \hline
Na 1&$5889.970$&$0.00$&$ 0.122$&$173.0\pm10.1$&$ 3.36$&$161.9\pm 8.7$&$ 2.83$&$191.8\pm11.8$&$ 3.50$&$ 87.6\pm 5.3$&$ 2.13$&$130.1\pm 8.3$&$ 2.96$\\ 
Na 1&$5895.924$&$0.00$&$-0.190$&$143.3\pm 8.2$&$ 3.25$&$122.1\pm 6.7$&$ 2.59$&$168.1\pm10.8$&$ 3.52$&$ 64.5\pm 4.4$&$ 2.12$&$102.3\pm 6.0$&$ 2.79$\\ 
\hline
Mg 1&$3832.304$&$2.71$&$ 0.150$&$167.1\pm11.8$&$ 4.46$&$202.7\pm16.6$&$ 4.58$&$204.8\pm11.6$&$ 4.68$&$129.0\pm10.0$&$ 3.92$&$           -$&$    -$\\ 
Mg 1&$3838.290$&$2.72$&$ 0.420$&$192.6\pm15.4$&$ 4.31$&$           -$&$    -$&$242.5\pm20.2$&$ 4.50$&$153.5\pm15.6$&$ 3.97$&$174.6\pm18.2$&$ 4.30$\\ 
Mg 1&$4167.271$&$4.35$&$-1.004$&$           -$&$    -$&$           -$&$    -$&$           -$&$    -$&$           -$&$    -$&$ 22.2\pm 3.4$&$ 4.92$\\ 
Mg 1&$5172.700$&$2.71$&$-0.390$&$197.2\pm11.3$&$ 4.81$&$190.0\pm10.2$&$ 4.49$&$211.0\pm11.4$&$ 4.85$&$119.7\pm 6.8$&$ 3.83$&$148.4\pm 8.4$&$ 4.46$\\
\hline
\hline
\end{tabular}
\end{table*}

\begin{table*}[h]
\centering
\scriptsize
\begin{tabular}{lrrr|rr|rr|rr|rr|rr} \hline \hline
El.&\multicolumn{1}{c}{$\lambda$}&$\chi_\mathrm{ex}$&$\log(gf)$&\multicolumn{10}{|c}{$EQW\pm\Delta EQWQ$~~[m\AA]~~~$\log\varepsilon(X)+12$~~[dex]} \\
&&&&\multicolumn{2}{|c}{ET0381}&\multicolumn{2}{|c}{scl002\_06}&\multicolumn{2}{|c}{scl\_03\_059}&\multicolumn{2}{|c}{scl031\_11}&\multicolumn{2}{|c}{scl074\_02}\\
 \hline \hline
Mn 1&$4030.753$&$0.00$&$-0.470$&$           -$&$    -$&$164.3\pm14.0$&$ 1.93$&$117.3\pm10.7$&$ 1.23$&$ 92.9\pm 6.9$&$ 0.95$&$112.7\pm 9.1$&$ 1.61$\\ 
Mn 1&$4033.062$&$0.00$&$-0.618$&$           -$&$    -$&$137.3\pm10.4$&$ 1.43$&$105.6\pm 7.5$&$ 1.25$&$ 81.6\pm 8.1$&$ 0.97$&$122.9\pm10.5$&$ 2.16$\\ 
Mn 1&$4034.483$&$0.00$&$-0.811$&$           -$&$    -$&$127.0\pm 9.4$&$ 1.50$&$110.0\pm 7.6$&$ 1.63$&$ 68.7\pm 9.4$&$ 1.00$&$109.7\pm10.6$&$ 2.08$\\ 
Mn 1&$4041.355$&$2.11$&$ 0.285$&$101.0\pm 6.6$&$ 3.01$&$           -$&$    -$&$           -$&$    -$&$           -$&$    -$&$           -$&$    -$\\ 
Mn 1&$4823.524$&$2.32$&$ 0.144$&$ 57.8\pm 5.4$&$ 2.68$&$ 23.8\pm 2.8$&$ 1.83$&$           -$&$    -$&$           -$&$    -$&$           -$&$    -$\\ 
\hline
Fe 1&$3753.611$&$2.18$&$-0.890$&$           -$&$    -$&$           -$&$    -$&$ 69.5\pm 8.6$&$ 4.13$&$           -$&$    -$&$           -$&$    -$\\
\hline
\hline
\end{tabular}
\end{table*}

\begin{table*}[h]
\begin{center}
\scriptsize
\begin{tabular}{lrrr|rr|rr|rr|rr|rr} \hline \hline
El.&\multicolumn{1}{c}{$\lambda$}&$\chi_\mathrm{ex}$&$\log(gf)$&\multicolumn{10}{|c}{$EQW\pm\Delta EQW$~~[m\AA]~~~$\log\varepsilon(X)+12$~~[dex]} \\
&&&&\multicolumn{2}{|c}{ET0381}&\multicolumn{2}{|c}{scl002\_06}&\multicolumn{2}{|c}{scl\_03\_059}&\multicolumn{2}{|c}{scl031\_11}&\multicolumn{2}{|c}{scl074\_02}\\
 \hline \hline
Fe 1&$5216.280$&$1.61$&$-2.102$&$113.8\pm 7.6$&$ 5.21$&$ 73.6\pm 5.1$&$ 4.04$&$ 74.6\pm 4.1$&$ 4.31$&$ 28.3\pm 2.4$&$ 3.72$&$ 46.8\pm 4.8$&$ 4.13$\\ 
Fe 1&$5232.950$&$2.94$&$-0.067$&$116.4\pm 7.1$&$ 4.90$&$           -$&$    -$&$ 79.4\pm 5.1$&$ 4.11$&$ 27.8\pm 2.8$&$ 3.33$&$ 56.9\pm 4.0$&$ 3.92$\\ 
Fe 1&$5266.555$&$3.00$&$-0.386$&$ 97.7\pm 5.5$&$ 4.94$&$ 53.1\pm 4.1$&$ 3.88$&$ 61.4\pm 3.3$&$ 4.17$&$           -$&$    -$&$ 53.6\pm 4.1$&$ 4.25$\\ 
Fe 1&$5281.790$&$3.04$&$-0.834$&$ 71.6\pm 4.2$&$ 4.95$&$ 27.7\pm 2.5$&$ 3.94$&$ 33.3\pm 2.1$&$ 4.16$&$           -$&$    -$&$           -$&$    -$\\ 
Fe 1&$5283.621$&$3.24$&$-0.520$&$ 88.1\pm 5.0$&$ 5.20$&$ 38.3\pm 3.6$&$ 4.09$&$ 42.5\pm 2.5$&$ 4.28$&$           -$&$    -$&$ 34.7\pm 2.6$&$ 4.34$\\ 
Fe 1&$5302.302$&$3.28$&$-0.880$&$ 57.9\pm 4.3$&$ 5.06$&$           -$&$    -$&$ 30.9\pm 3.8$&$ 4.47$&$           -$&$    -$&$           -$&$    -$\\
\hline
\hline
\end{tabular}
\end{center}
\noindent $^1$ Assuming $f_\mathrm{odd}=0.18$, LTE, and plane-parallel transfer, and correcting for \ion{Fe}{i} blends.\\
\noindent $^2$ From synthesis
\end{table*}

\begin{table*}[h]
\caption{Magnitudes, CaT metallicities, and corresponding photometric effective temperatures
and surface gravities of our stars.}
\label{table:photometry}
\centering
\begin{tabular}{lrrrrcrrrrr} \hline \hline
\multicolumn{1}{c}{ID}&\multicolumn{1}{c}{V$\pm\sigma_\mathrm{V}$}&\multicolumn{1}{c}{I$\pm\sigma_\mathrm{I}$}&
\multicolumn{1}{c}{J}&\multicolumn{1}{c}{K}&\multicolumn{1}{c}{[Fe/H]$_\mathrm{CaT}$}&
\multicolumn{4}{c}{Photometric~$T_\mathrm{eff}$~[K]}&\multicolumn{1}{c}{$\log g$}  \\
&&&&&&\multicolumn{1}{c}{V--I}&\multicolumn{1}{c}{V--J}&\multicolumn{1}{c}{V--K}&\multicolumn{1}{c}{mean}&
\multicolumn{1}{c}{[cgs]}\\ \hline
ET0381       & $18.04$  & $16.92$ &$16.09$ & $15.44$ & $-2.75$ & $4586$ & $4551$ & $4559$ & $4565$ & $1.17$ \\
scl002\_06   & $17.12$  & $15.78$ &$14.90$ & $14.34$ & $-3.04$ & $4351$ & $4359$ & $4469$ & $4393$ & $0.69$ \\
scl\_03\_059 & $17.93$  & $16.74$ &$15.93$ & $15.37$ & $-2.82$ & $4490$ & $4520$ & $4592$ & $4534$ & $1.10$ \\
scl031\_11   & $17.80$  & $16.65$ &$15.88$ & $15.21$ & $-3.61$ & $4650$ & $4729$ & $4627$ & $4669$ & $1.12$ \\
scl074\_02   & $18.06$  & $16.96$ &$16.15$ & $15.69$ & $-3.02$ & $4645$ & $4637$ & $4764$ & $4682$ & $1.23$ \\ \hline
\end{tabular}
\end{table*}

\begin{table*}[h!]
\caption{Final spectroscopic stellar parameters}
\label{table:atmo}
\centering
\begin{tabular}{lrrrr} \hline \hline
\multicolumn{1}{c}{ID}&\multicolumn{1}{c}{$T_\mathrm{eff}$}&\multicolumn{1}{c}{$\log g$}&
\multicolumn{1}{c}{[Fe/H]}&\multicolumn{1}{c}{$v_\mathrm{t}$} \\
&\multicolumn{1}{c}{[K]}&\multicolumn{1}{c}{[cgs]}&model&\multicolumn{1}{c}{[km\,s$^{-1}$]} \\ \hline
ET0381      &$4540$&$1.15$&$-2.45$&$1.70$ \\
scl002\_06  &$4300$&$0.63$&$-3.45$&$2.15$ \\
scl\_03\_059&$4400$&$1.01$&$-3.20$&$1.78$ \\
scl031\_11  &$4550$&$1.05$&$-3.88$&$2.10$ \\
scl074\_02  &$4600$&$1.19$&$-3.30$&$1.80$ \\ \hline
\end{tabular}
\end{table*}

\begin{table*}[h]
\caption{Changes in the microturbulence velocity and in the mean abundances [X/H]
caused by a $100$~K increase in $T_\mathrm{eff}$, accompanied by the corresponding
$\log g$ change according to Eq.~\ref{eq:logg}. The first two lines give the change
for the carbon when \Teff is increased or decreased by $100$~K, because
the change is strongly asymmetric.}
\label{table:Tp_del_abun}
\centering
\begin{tabular}{l|rrrrr} \hline \hline
                        &ET0381 &scl002\_06&scl\_03\_059&scl031\_11&scl074\_02 \\ \hline \hline
$\Delta v_\mathrm{t}$ (km\,s$^{-1}$)& 0.18  &   0.18   &  0.20  &   0.03   &  0.15\\ \hline
                        &\multicolumn{5}{c}{$\Delta$[X/H] (dex)}      \\ \hline
C $+$                   &$ 0.20$  & $  0.40$   & $ 0.30$  &  $ 0.35$   & $ 0.25$\\
C $-$                   &$-0.20$  & $ -0.10$   & $-0.20$  &  $-0.15$   & $-0.15$\\ \hline
Na  I                   &$ 0.10$  & $  0.08$   & $ 0.09$  &  $ 0.10$   & $ 0.07$\\
Mg  I                   &$ 0.09$  & $  0.06$   & $ 0.10$  &  $ 0.07$   & $ 0.07$\\
Al  I                   &$ 0.11$  & $  0.10$   & $ 0.08$  &  $ 0.10$   & $ 0.06$\\
Si  I                   &$ 0.07$  & $  0.07$   & $ 0.09$  &  $ 0.11$   & $ 0.12$\\
Ca  I                   &$ 0.06$  & $  0.07$   & $ 0.06$  &  $ 0.07$   & $ 0.06$\\
Sc II                   &$ 0.04$  & $  0.05$   & $ 0.03$  &  $ 0.08$   & $ 0.04$\\
Ti  I                   &$ 0.16$  & $  0.18$   & $ 0.15$  &  $ 0.13$   & $ 0.12$\\
Ti II                   &$ 0.02$  & $  0.03$   & $ 0.02$  &  $ 0.07$   & $ 0.04$\\
Cr  I                   &$ 0.10$  & $  0.13$   & $ 0.11$  &  $ 0.11$   & $ 0.10$\\
Mn  I                   &$ 0.07$  & $  0.10$   & $ 0.10$  &  $ 0.13$   & $ 0.12$\\
Fe  I                   &$ 0.08$  & $  0.09$   & $ 0.08$  &  $ 0.10$   & $ 0.09$\\
Fe II                   &$-0.01$  & $ -0.00$   & $-0.03$  &  $ 0.02$   & $ 0.01$\\
Co  I                   &$ 0.08$  & $  0.12$   & $ 0.07$  &  $ 0.15$   & $ 0.12$\\
Ni  I                   &$ 0.07$  & $  0.09$   & $ 0.08$  &  $ 0.11$   & $ 0.10$\\
Zn  I                   &$ 0.03$  & $  0.03$   & $ -   $  &  $ -   $   & $ 0.04$\\
Sr II                   &$ 0.03$  & $ -0.04$   & $ 0.00$  &  $ 0.09$   & $ 0.02$\\
Ba II                   &$ 0.07$  & $  -   $   & $ 0.05$  &  $ -   $   & $ -   $\\ \hline
\end{tabular}
\end{table*}

\begin{table*} 
 \caption{\label{Tab:hfs} Barium LTE and NLTE abundances, NLTE abundance corrections, and the effect of uncertainties in stellar parameters on abundances (in dex) derived from individual lines of \ion{Ba}{ii} in ET0381 and scl\_03\_059.}
\scriptsize
\centering
\begin{tabular}{lrrrrrrcrrrrrrr}
\hline\hline \noalign{\smallskip}
 \ Object & \multicolumn{6}{c}{ET0381} & & \multicolumn{7}{c}{scl\_03\_059} \\
\noalign{\smallskip} \cline{2-7} \cline{9-15} \noalign{\smallskip}
 \ $\lambda$, \AA & \multicolumn{4}{c}{4934} & 6141 & 6496 & & \multicolumn{4}{c}{4934} & 5853 & 6141 & 6496 \\
\noalign{\smallskip} \cline{2-5} \cline{9-12} \noalign{\smallskip}
 \ $f_\mathrm{odd}$ & 0.72 & 0.46 & 0.18 & 0.11 & 0.46 & 0.46 & & 0.72 & 0.46 & 0.18 & 0.11 & 0.46 & 0.46 & 0.46 \\
\hline \noalign{\smallskip}
$\eps{}$, LTE & $-$1.82 & $-$1.74 & $-$1.62 & $-$1.56 & $-$1.52 & $-$1.54 & & $-$1.72 & $-$1.58 & $-$1.40 & $-$1.30 & $-$1.29 & $-$1.41 & $-$1.39 \\
\noalign{\smallskip} \cline{6-7} \cline{13-15} \noalign{\smallskip}
    & & & & & \multicolumn{2}{l}{mean: $-$1.53{\scriptsize $\pm$0.02}} & & & & & & \multicolumn{3}{l}{mean: $-$1.36{\scriptsize $\pm$0.05}} \\
$\Delta_{\rm NLTE}$ & 0.01 & 0.01 & 0.00 & 0.00 & 0.08 & 0.04 & & 0.00 & 0.00 & 0.00& $-$0.01 & 0.08 & 0.10 & 0.05 \\
$\eps{}$, NLTE & $-$1.81 & $-$1.72 & $-$1.62 & $-$1.56 & $-$1.44 & $-$1.50 & & $-$1.72 & $-$1.58 & $-$1.40 & $-$1.31 & $-$1.21 & $-$1.31 & $-$1.34 \\
\noalign{\smallskip} \cline{6-7} \cline{13-15} \noalign{\smallskip}
    & & & & & \multicolumn{2}{l}{mean: $-$1.47{\scriptsize $\pm$0.04}} & & & & & & \multicolumn{3}{l}{mean: $-$1.29{\scriptsize $\pm$0.05}} \\
\end{tabular}
\end{table*}

\begin{landscape}
\begin{table}[h]
\begin{center}
\caption{Abundances}
\tiny
\tabcolsep 4pt
\label{tab:abundances}
\begin{tabular}{lrrrrrrrrrrrrrrrrrrrr}
\hline
         &  FeI  &  FeII &    C   &  NaI &   MgI &   AlI &  SiI  &  CaI  & ScII  &  TiI  &  TiII &  CrI  &   MnI &  CoI  &  NiI  &  Zn   &  SrII &   YII &  BaII &  EuII  \\
         &       &       &        &      &       &       &       &       &       &       &       &       &       &       &       &       &       &       &       &        \\      

\textbf{log$\epsilon$(X)$_{\odot}$}
         &   7.50&   7.50&    8.55&  6.33&   7.58&   6.47&   7.55&   6.36&   3.17&   5.02&   5.02&   5.67&   5.39&   4.92&   6.25&   4.60&   2.97&   2.24&   2.13&    0.51\\              
\hline     
 &&&&&&&&&&&&&&&&&&&& \\   
\textbf{ET0381}     &&&&&&&&&&&&&&&&&&&& \\   
&&&&&&&&&&&&&&&&&&&& \\   

Nb Lines &   80  & 11    &  1     &   2   &  5    & 1     & 2     &  9   &  5    &  10   & 24    &    6  &    3  &     4 &    3  &   1   &    1  &    2  &    3  &  1     \\    
$[$X/H]  &$-$2.44&$-$2.13&$-$3.44&$-$3.03&$-$2.91&$-$3.55&$-$2.81&$-$2.67&$-$2.87&$-$2.97&$-$2.56&$-$2.75&$-$2.63&$-$2.94&$-$2.65&$<-2.77$&$-$3.30&$<-3.31$&$-$3.64&$<-$2.81\\
\end{tabular}
\end{center}
\end{table}
\end{landscape}

\end{document}